\documentclass[aps,pra,twocolumn]{revtex4-1}
\usepackage{amsmath,amssymb}
\usepackage{natbib}
\usepackage{braket}
\usepackage{hyperref}
\usepackage{graphicx}
\usepackage{color,soul}
\usepackage[caption=false]{subfig}
\usepackage{epstopdf}
\DeclareMathOperator{\Tr}{Tr}
\newcommand{\RN}[1]{%
  \textup{\uppercase\expandafter{\romannumeral#1}}%
}
    \newcommand{\Eq}[1]		{Eq.~(\ref{#1})}		
    \newcommand{\Eqs}[2]	{Eqs.~(\ref{#1}-\ref{#2})}	
    \newcommand{\Fig}[1]	{Fig.~\ref{#1}}			

\begin{document}
\title{Cutoff-free Circuit Quantum Electrodynamics}
\author{Moein Malekakhlagh}
\author{Alexandru Petrescu}
\author{Hakan E. T\"ureci}
\affiliation{Department of Electrical Engineering, Princeton University, Princeton, New Jersey, 08544}
\date{\today}
\begin{abstract}
Any quantum-confined electronic system coupled to the electromagnetic continuum is subject to radiative decay and renormalization of its energy levels. When coupled to a cavity, these quantities can be strongly modified with respect to their values in vacuum. Generally, this modification can be accurately captured by including only the closest resonant mode of the cavity. In the circuit quantum electrodynamics architecture, it is however found that the radiative decay rates are strongly influenced by far off-resonant modes. A multimode calculation accounting for the infinite set of cavity modes leads to divergences unless a cutoff is imposed. It has so far not been identified what the source of divergence is. We show here that unless gauge invariance is respected, any attempt at the calculation of circuit QED quantities is bound to diverge. We then present a theoretical approach to the calculation of a finite spontaneous emission rate and the Lamb shift that is free of cutoff.
\end{abstract}
\maketitle

\textit{Introduction.} An atom-like degree of freedom coupled to continuum of electromagnetic (EM) modes spontaneously decays. When the atom is confined in a resonator, the emission rate can be modified compared with its value in free space, depending on the EM local density of states at the atomic position \cite{Kleppner_Inhibited_1981, Goy_Observation_1983, Hulet_Inhibited_1985, Jhe_Suppression_1987}, which is called the Purcell effect \cite{Purcell_Resonance_1946}. An accompanying effect is the Lamb shift, a radiative level shift first observed in the microwave spectroscopy of the hydrogen $^2P_{1/2}-~^2S_{1/2}$ transition \cite{Lamb_Fine_1947}. These quantities have been experimentally accurately characterized for superconducting Josephson junction (JJ) based qubits coupled to coplanar transmission lines \cite{Fragner_Resolving_2008, Houck_Controlling_2008} and three-dimensional resonators \cite{Nigg_BlackBox_2012}. In the dispersive regime where a qubit with transition frequency $\omega_j$ is far-detuned from the nearest resonant cavity mode (frequency $\nu_r$, loss $\kappa_r$), single mode expressions exist for the Purcell decay rate, $\gamma_P = (g/\delta)^2 \kappa_r$ and the Lamb shift, $\Delta_L = g^2 / \delta$. Here $g$ denotes the coupling between the qubit and the cavity mode and $\delta=\omega_j - \nu_r$ denotes their detuning \cite{Boissonneault_Dispersive_2009}. However, for large couplings accessible in circuit QED, the single mode approximation is often inaccurate \cite{Fragner_Resolving_2008, Houck_Controlling_2008}. In addition, due to particular boundary conditions imposed by the capacitive coupling of a resonator to external waveguides, the qubit relaxation time is limited by the EM modes that are far-detuned from the qubit frequency \cite{Houck_Controlling_2008}. Similarly the measured Lamb shift in the dispersive regime can only be accurately fit with an extended Jaynes-Cummings (JC) model including several modes and qubit levels \cite{Fragner_Resolving_2008}. The Purcell rate has been generalized to account for all modes
\begin{figure}[t!]
\subfloat[\label{Fig:cQED-open}]{%
\includegraphics[scale=0.45]{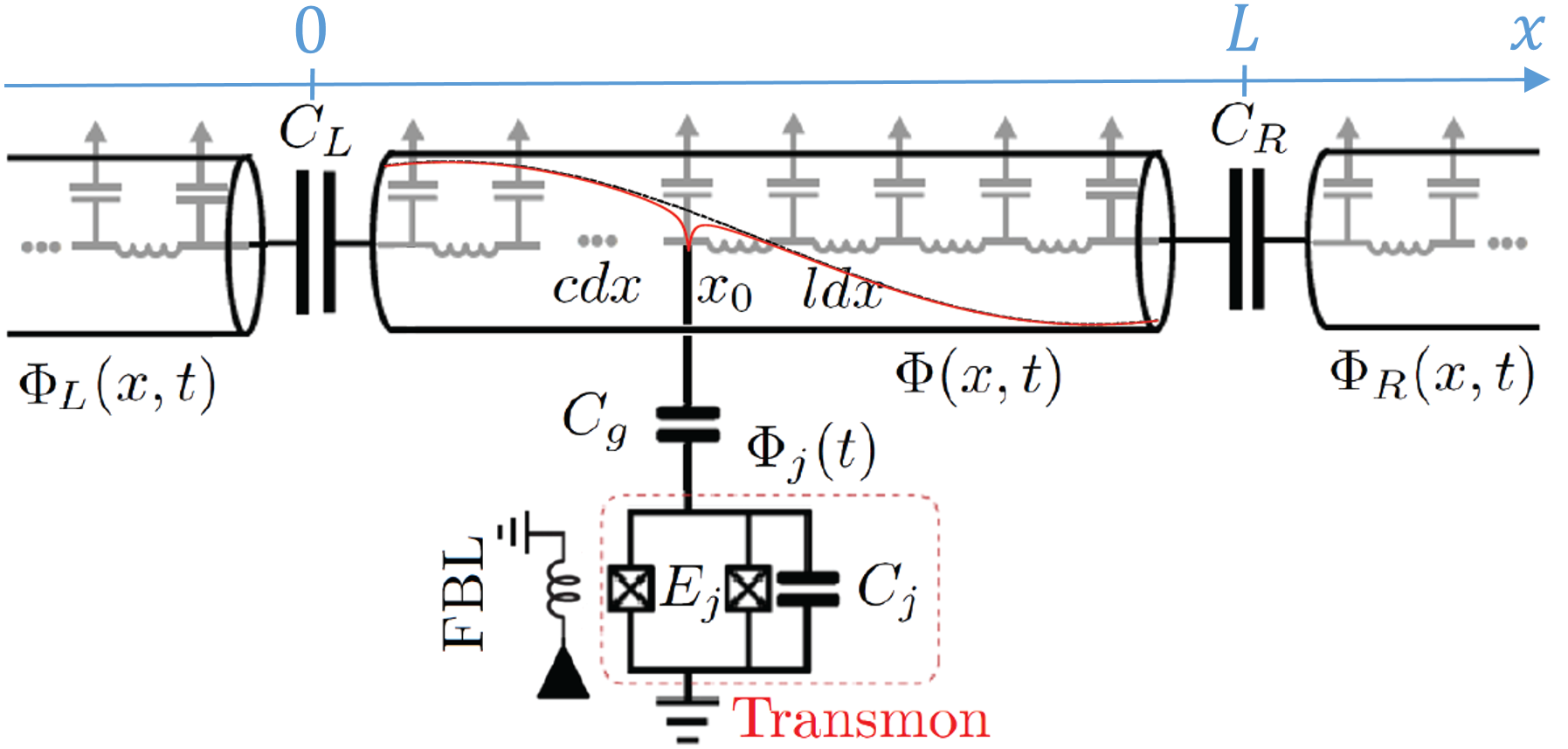}%
}
\hfill
\subfloat[\label{Fig:poles_sf_labeled}]{%
\includegraphics[width=\columnwidth]{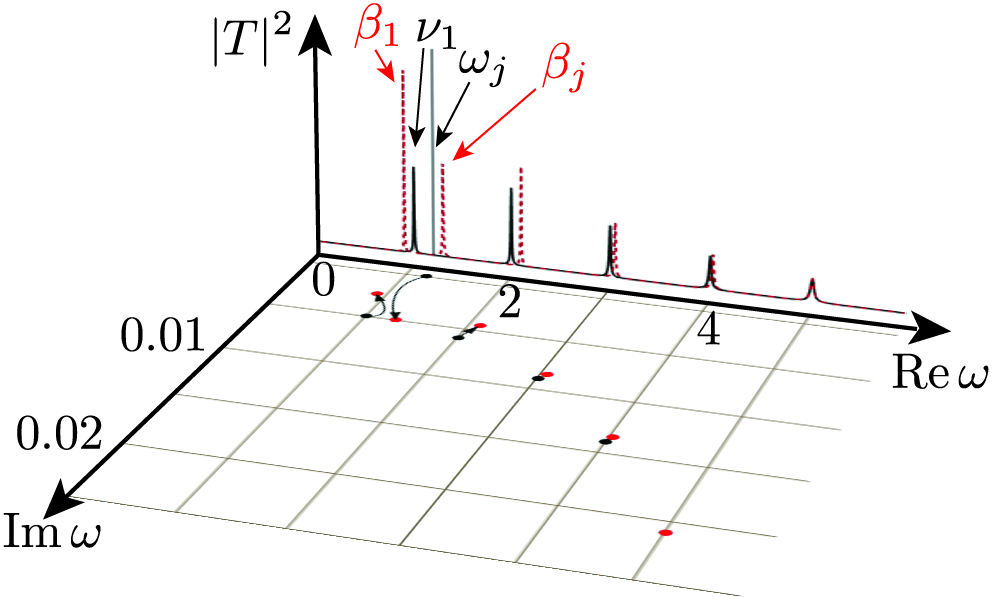}%
}
\caption{a) A transmon qubit coupled to an open superconducting resonator. The black dashed line is a cartoon of the fundamental bare mode of the resonator, while the red solid curve represents the modified resonator mode. b) The transmission  $|T|^2$ is shown versus the real frequency for the bare resonator modes (solid black curves). Capacitively coupling the qubit, whose transition frequency $\omega_j$ is slightly above the fundamental resonator frequency $\nu_1$, gives rise to hybridized modes (dashed red curves). Alternatively, one may study the positions of these resonances in the complex frequency plane, where the bare resonator and qubit poles (black points) are displaced into hybridized resonator-like and qubit-like resonances (red points). The Purcell decay and the Lamb shift are obtained as the displacement of the qubit-like pole. The bare (hybridized) complex frequencies are the poles (zeros) of the characteristic function $D_j(s)$.} 
\end{figure}
\begin{equation}
\gamma_P=\sum_n(g_n/\delta_n)^2 \, \kappa_n,
\label{eqn:Multimode Purcell Rate} 
\end{equation}
where $g_n$ and $\delta_n =\omega_j - \nu_n$ are coupling to and detuning from resonator mode $n$ with frequency $\nu_n$ and decay rate $\kappa_n$. Expression~(\ref{eqn:Multimode Purcell Rate}) is divergent without imposing a high-frequency cutoff \cite{Houck_Controlling_2008}. Divergences appear as well in the Lamb shift and other vacuum-induced phenomena, e.g. photon-mediated qubit-qubit interactions \cite{Filipp_Multimode_2011}. These divergences are neither specific to the dispersive limit nor to the calculational scheme used to compute QED quantities. This issue is well-known for the Lamb shift \cite{Lamb_Fine_1947}, but less noted for the spontaneous emission rate. Indeed, free space spontaneous emission rate diverges as well, as we show in \cite{Supplementary}. The finite result by Wigner and Weisskopf \cite{Weisskopf_Berechnung_1930, Scully_Quantum_1997} is due to Markov approximation which filters out the ultraviolet divergence. Recent generalizations of the Wigner-Weisskopf approach impose an artificial cut-off to obtain a finite result \cite{Krimer_Route_2014}. So far, no satisfactory theoretical explanation has been given for these divergences. Here we address this issue within the framework of circuit quantum electrodynamics~\cite{Devoret_Quantum_1995} (QED) and show that finite expressions can be obtained when gauge invariance is respected. We focus here on a superconducting artificial atom coupled to an open transmission-line resonator, but our results should be valid for other types of one-dimensional open EM environments as well. 


\textit{Gauge invariance in circuit QED.} The role of gauge invariance in accounting for light-matter interaction has been a vexing question since the beginnings of QED (see Ref.~\cite{Lamb_Matter-field_1987}, and references therein). Hence, we first discuss gauge invariance in  superconducting electrical circuits, and its impact on QED observables.  

We consider a weakly nonlinear charge qubit (e.g. transmon \cite{Koch_Charge_2007}\cite{Schreier_Suppressing_2008}) capacitively coupled to a transmission-line resonator that in turn is coupled at both ends to semi-infinite waveguides (\Fig{Fig:cQED-open}). We assign flux variables to nodes, $\Phi_n (t) = \int^t d\tau \, V_n(\tau)$, with $V_n(t)$ being the instantaneous voltage at node $n$ with respect to the ground node \cite{Devoret_Quantum_1995, Devoret_Quantum_2014}. Fixing the ground amounts to a particular gauge choice \cite{Devoret_Quantum_1995}. For the connection geometry in Fig~\ref{Fig:cQED-open}, the light-matter interaction derives from the energy on the coupling capacitor in the dipole approximation, $T_\text{int} = \frac{1}{2} C_g [\dot{\Phi}(x_0) - \dot{\Phi}_j]^2$ \cite{Supplementary}, with $x_0$ the qubit position. If from the three terms in its expansion, $T_\text{EM} =  \frac{1}{2} C_g  \dot{\Phi}(x_0)^2$, $T_ \text{EM-JJ} = - C_g \dot{\Phi}(x_0) \cdot \dot{\Phi}_j$ and $T_\text{JJ} = \frac{1}{2} C_g  \dot{\Phi}_j^2$, only the direct interaction $T_ \text{EM-JJ}$ is kept, a multimode JC model in terms of circuit parameters can be derived \footnote{See also the Supplementary Material for a brief derivation of the Heisenberg-Langevin equation of motion and a discussion of the multimode convergence of its characteristic function, which includes Refs.~\cite{Scully_Quantum_1997, Bishop_Circuit_2010, Devoret_Quantum_2014, Malekakhlagh_NonMarkovian_2016, Malekakhlagh_Origin_2016, Clerk_Introduction_2010, Morse_Methods_1953, Economou_Green_1984, Hassani_Mathematical_2013, Tureci_SelfConsistent_2006, Boissonneault_Dispersive_2009, Senitzky_Dissipation_1960}}, but gives rise to a diverging Purcell rate using Eq.~(\ref{eqn:Multimode Purcell Rate}). This open JC Model involves a two level approximation (TLA) of the JJ Hilbert space, the rotating wave approximation (RWA) to drop nonresonant contributions, and the Born and Markov approximations leading to a Master equation accounting for losses due to resonator-waveguide coupling. It is unclear which approximation underlies the divergence, or whether the divergence can be resolved within the effective subgap circuit QED field theory.

We first note that keeping only the direct interaction $T_ \text{EM-JJ}$ violates gauge invariance. We find that inclusion of all terms, in particular $T_\text{EM}$, equivalent to the diamagnetic $A^2$ term in the minimal coupling Hamiltonian $(p - e A)^2/2m$ \cite{Malekakhlagh_Origin_2016}, is essential to make all studied QED observables finite. 

The $A^2$-term is thought to have no impact on transition frequencies in vacuum-induced effects such as the Lamb shift. Because it does not involve atomic operators, it is expected to make the same perturbative contribution to every atomic energy level, precluding observable shifts in {\it transition} frequencies \cite{Milonni_Quantum_2013}. This argument relies on perturbation theory in the $A^2$-term. We show that the diamagnetic term {\it does} have an impact when accounted for exactly to all orders.  

\textit{Heisenberg equations of motion} describing the infinite network in \Fig{Fig:cQED-open}, extending from $x=-\infty$ to $x=\infty$, are \cite{Malekakhlagh_NonMarkovian_2016, Supplementary}
\begin{eqnarray}
&\hat{\ddot{\varphi}}_j(t)+ ( 1- \gamma) \omega_j^2\sin{[\hat{\varphi}_j(t)]}  = \gamma\partial_{t}^2\hat{\varphi}(x_0,t),  \label{eqn:TransDyn} \\
&\left[\partial_{x}^2-\chi(x,x_0)\partial_{t}^2\right]\hat{\varphi}(x,t)  = \chi_s\omega_j^2 \sin{[\hat{\varphi}_j(t)]}\delta(x-x_0),
\label{eqn:ResDyn}
\end{eqnarray}
Here $\hat{\varphi}_j(t)$ and $\hat{\varphi}(x,t)$ are dimensionless flux operators for the JJ and the resonator-waveguide system, respectively, $\gamma \equiv C_g/(C_g+C_j)$ is a capacitive  ratio, $\chi_s = \gamma C_j / cL$ is the dimensionless series capacitance of $C_g$ and $C_j$, $\omega_j$ is the dimensionless transmon frequency, and $\chi_i\equiv C_i/(cL)$ for $i=g,j,R,L$ \cite{Supplementary}. These two inhomogeneous equations show that the flux field at $x_0$ drives the dynamics of the JJ [\Eq{eqn:TransDyn}], while the JJ acts as a source driving the EM fields [\Eq{eqn:ResDyn}]. In addition, the fields are subject to continuity conditions at the ends of the resonator $x=0,1$ (in units of $L$).

It is instructive to trace the individual terms of $T_\text{int}$ in Eqs.~(\ref{eqn:TransDyn}-\ref{eqn:ResDyn}). $T_\text{JJ}$ modifies the qubit frequency, renormalizing $\gamma$ from $C_g/C_j$ to $C_g/(C_g+C_j)$, while the direct interaction term  $T_ \text{EM-JJ}$ gives source terms in both equations. Most importantly, $T_\text{EM}$ introduces an effective scattering term in the wave equation describing the fields in the transmission line, by modifying the unitless capacitance per length from $1$ to $\chi(x,x_0) = 1 + \chi_s \delta(x-x_0)$. Consequently, these equations are consistent \cite{Malekakhlagh_Origin_2016} with Kirchhoff's law of current conservation. 
In particular, at $x=x_0$, Eq.~(\ref{eqn:ResDyn}) yields $\left.\partial_x\hat{\varphi}(x,t)\right]_{x_0^-}^{x_0^+}=\chi_s\partial_t^2\hat{\varphi}(x_0,t)+\chi_s\omega_j^2\sin[\hat{\varphi}_j(t)]$, where the discontinuity in the resonator current is equal to the total current through the capacitive and Josephson branches of the transmon. Similar modification of resonator dynamics has been pointed out before for JJ-based qubits \cite{Nigg_BlackBox_2012, Bourassa_Josephson_2012, Malekakhlagh_Origin_2016}.

Equation~\ref{eqn:ResDyn} can be solved in the Fourier domain, where $\hat{\tilde{\varphi}}(x,\omega) = \int_{-\infty}^{\infty} dt \, \hat{\varphi}(x,t) e^{-i\omega t}$ can be expanded in the basis $\tilde{\varphi}_n (x,\omega)$ that solves the generalized eigenvalue problem $\left[\partial_x^2 + \chi(x,x_0) \, \omega^2 \right] \tilde{\varphi}_n (x,\omega)=0$, subject to continuity conditions at the ends of the resonator, i.e. $\partial_x \tilde{\varphi}_n (1^-,\omega)= \chi_{R} \omega^2[ \tilde{\varphi}_n (1^-,\omega)-\tilde{\varphi}_n (1^+,\omega)]$ and  $\partial_x \tilde{\varphi}_n (0^+,\omega)= \chi_{L} \omega^2 [\tilde{\varphi}_n (0^-,\omega)-\tilde{\varphi}_n (0^+,\omega)]$, which models the coupling to the waveguides and associated loss. The Dirac $\delta$-function in $\chi(x,x_0)$ leads to the discontinuity
\begin{equation}
-\left.\partial_x \tilde{\varphi}_n(x)\right]_{x_0^-}^{x_0^+} = \chi_s \omega_n^2 \tilde{\varphi}_n(x_0),
\label{eqn:Current-Conservation} 
\end{equation}
resulting in a modified current-conserving (CC) basis \cite{Malekakhlagh_Origin_2016}. These modifications in the spectrum of the transmission line resonator impact the qubit dynamics that is driven by resonator fluctuations.

\begin{figure}[t!]
\centering
\subfloat[\label{subfig:gXrXl1Em3}]{%
\includegraphics[scale=0.380]{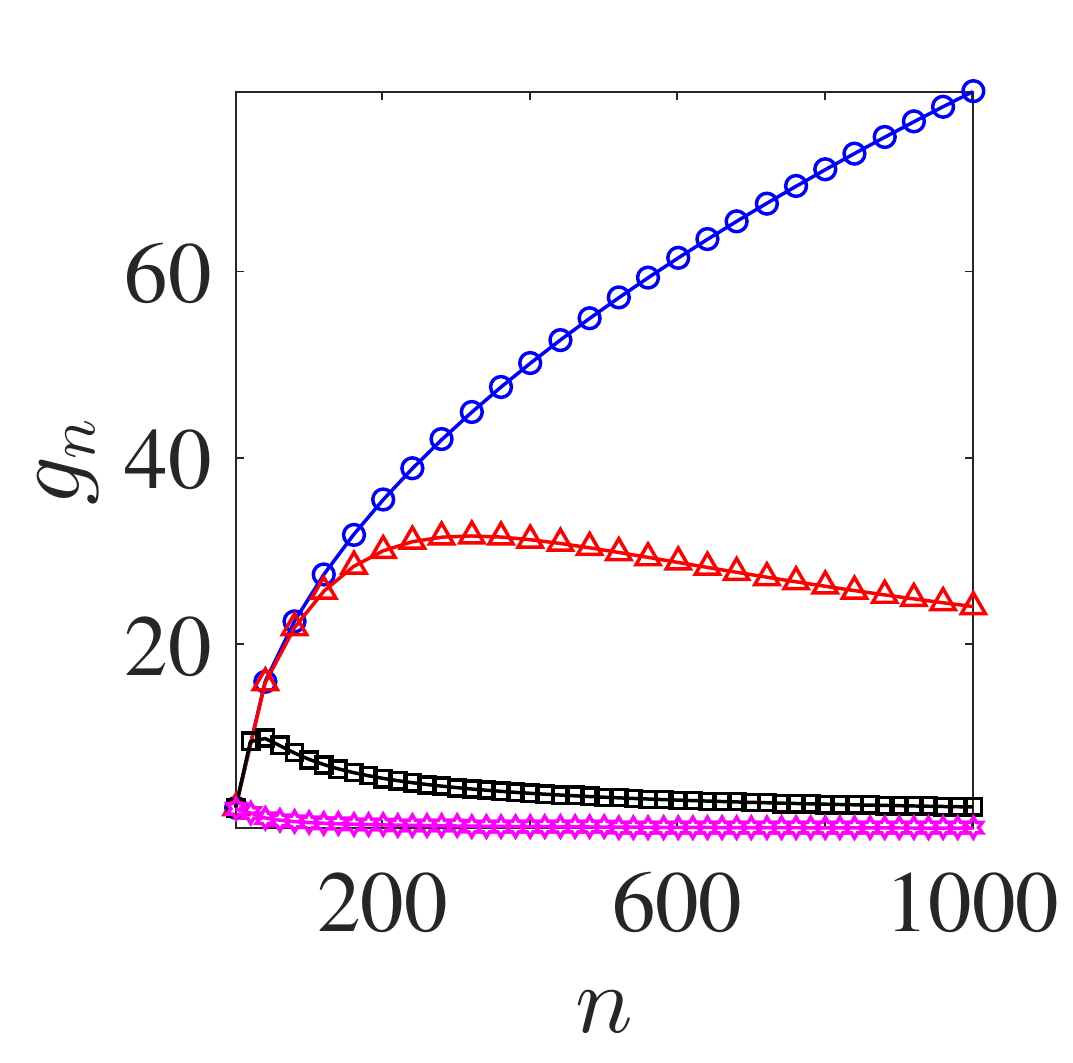}%
}\hfill
\subfloat[\label{subfig:KappaXrXl1Em3}]{%
\includegraphics[scale=0.380]{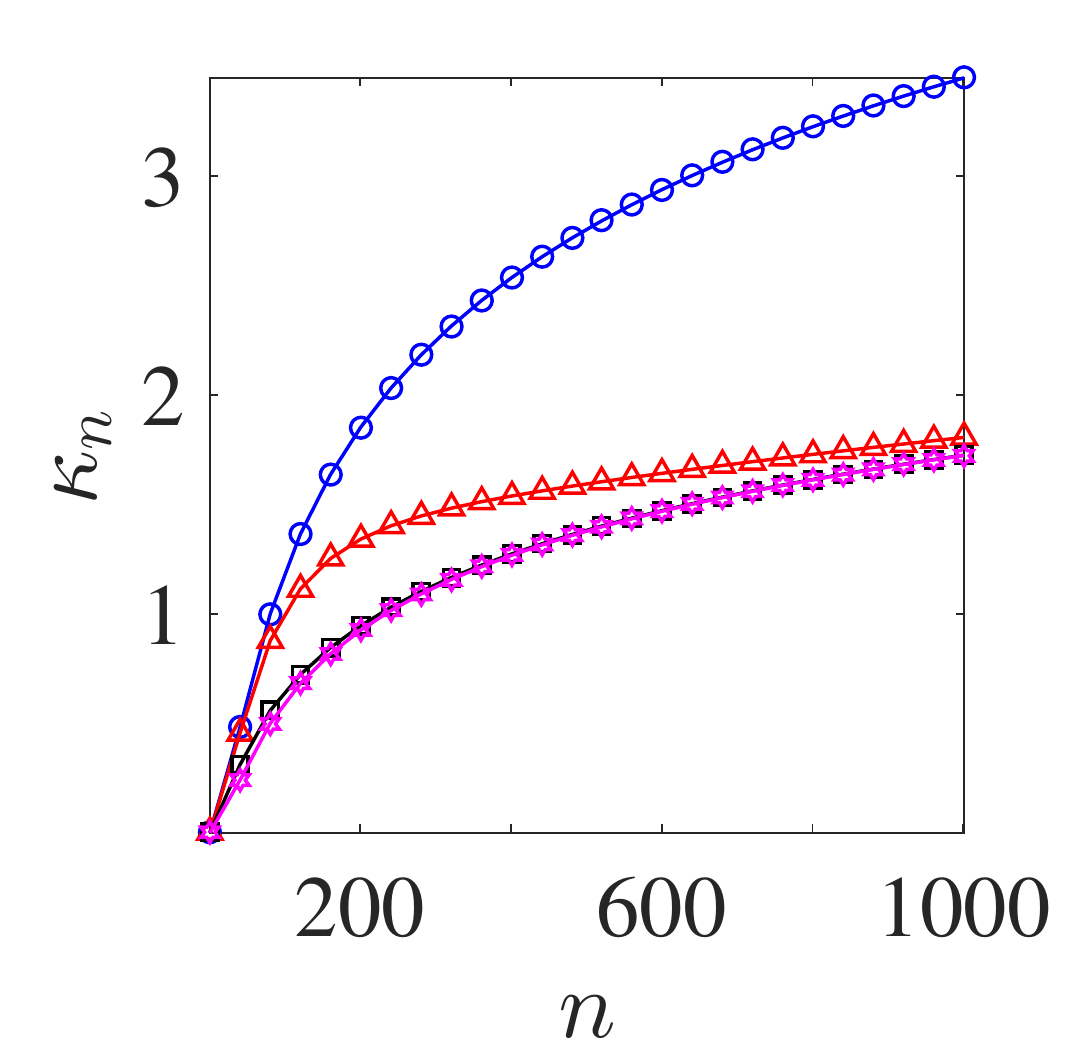}%
}\hfill
\subfloat[\label{subfig:LogLogSpEmRateXrXl1Em3}]{%
\includegraphics[scale=0.27]{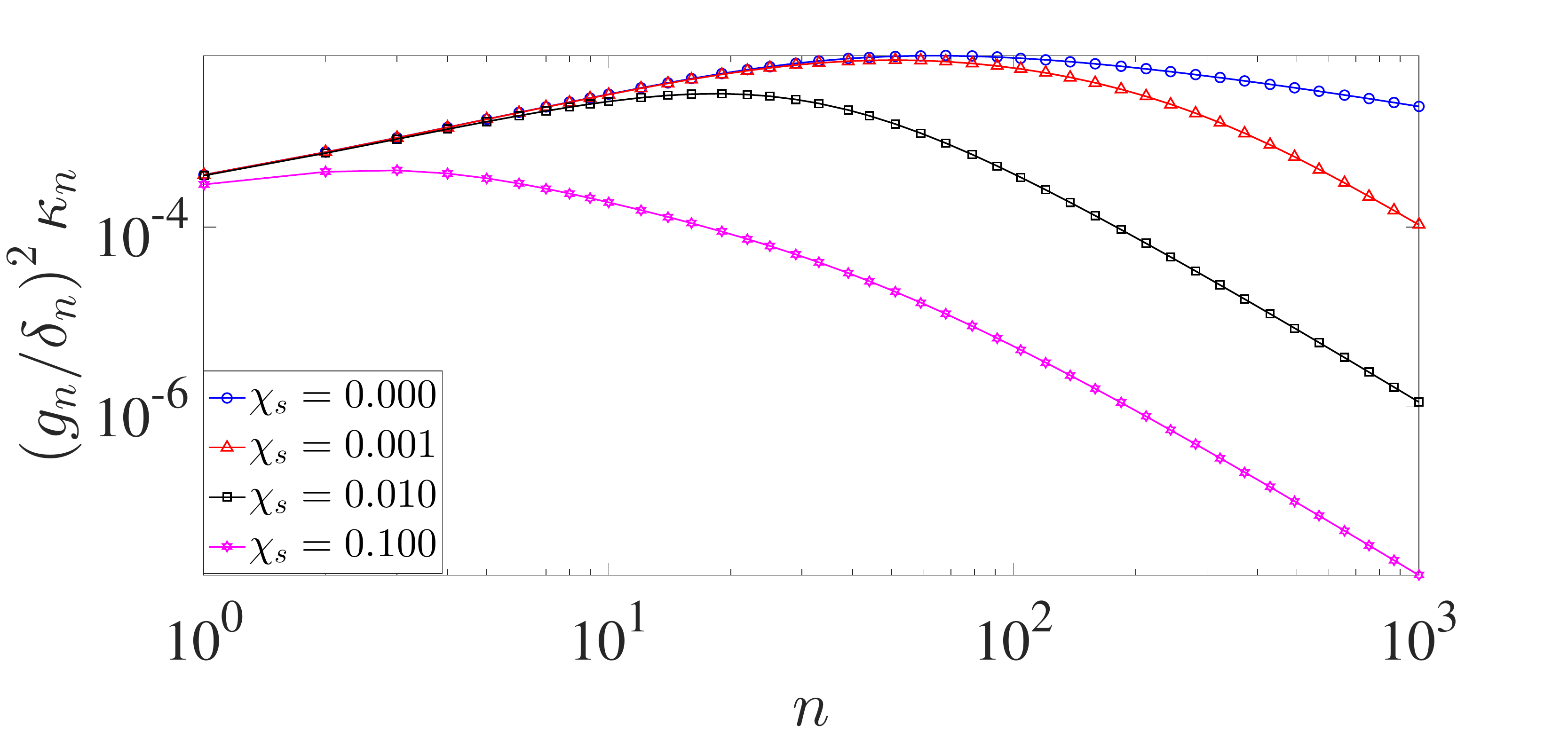}%
}
\caption{(Color online) Dependence of a) coupling strength $g_n$, b) resonator decay rate $\kappa_n$ (See \cite{Supplementary} for derivation) and c) Purcell decay rate in the dispersive regime $(g_n/\delta_n)^2\kappa_n$ on mode number $n$ for different values of $\chi_s=\{0,10^{-3},10^{-2},10^{-1}\}$. Other parameters are set as $\chi_R=\chi_L=10^{-3}$ and $x_0=0^+$.} 
\label{Fig:Effect of nonzero Xs}
\end{figure}

The role of modal modification in \Eq{eqn:Current-Conservation} can be illustrated with a phenomenological model. Previously, the Purcell rate and the Lamb shift have been calculated using the Lindblad formalism in the dispersive limit \cite{Boissonneault_Dispersive_2009}. An effective multimode JC model
\begin{equation}
\hat{\mathcal{H}}_\text{JC} = \frac{\omega_j}{2} \hat{\sigma}_z  + \sum_n \nu_n \hat{a}_n^\dagger \hat{a}_n + \sum_n g_n \left(\hat{\sigma}^+\hat{a}_n+\hat{\sigma}^- \hat{a}_n^{\dag}\right) 
\end{equation}
can be obtained from our first principles model \cite{Supplementary}, which incorporates the modifications to the resonator modes and the qubit dynamics. Resonator losses are included through a Bloch--Redfield equivalent zero-temperature master equation for the reduced density matrix of the resonator and qubit $\hat{\dot{\rho}} = -i [ \hat{\mathcal{H}}_\text{JC}, \hat{\rho} ] + \kappa_n \left( 2 \hat{a}_n \hat{\rho} \hat{a}_n^\dagger - \{ \hat{\rho}, \hat{a}_n^\dagger \hat{a}_n \} \right)$. The expressions of cavity frequencies $\nu_n$, associated losses $\kappa_n$ and modal interaction strengths $g_n$ are given in the Supplementary Material \cite{Supplementary}. All these quantities are functions of $\chi_s$, the strength of the modification of the capacitance per unit length. In particular, the light-matter coupling is found as $g_n=\frac{1}{2}\gamma\sqrt{\chi_j}\sqrt{\omega_j\nu_n}\tilde{\varphi}_n(x_0)$. We show in Fig.~\ref{subfig:gXrXl1Em3} that $g_n$ is non-monotonic \cite{Malekakhlagh_Origin_2016} for any $\chi_s\neq 0$, first increasing, then turning over at a critical $\chi_s$-dependent mode $n$, decreasing as $g_n \sim 1/\sqrt{n}$ in the large-$n$ limit \cite{Supplementary}. This high frequency behavior of $g_n$ renders the multimode Purcell rate finite, without an imposed cutoff \footnote{We note that this result is valid in the dispersive limit i.e. away from cavity resonances. In that limit, we expect this result to be fairly accurate when compared to the rate extracted from the exact time evolution of the Master equation for the multimode JC model.}.  

This phenomenon is not specific to the resonator geometry in \Fig{Fig:cQED-open}. The underlying physics is the conservation of current at the position $x_0$ of the qubit. At high frequency, the series capacitance $\chi_s$ becomes a short-circuit to ground, acting as a low-pass filter and suppressing mode amplitude at $x_0$. This is the cause of the power law drop of $g_n$ as $n \rightarrow \infty$ (\Fig{subfig:gXrXl1Em3}). Moreover, eliminating the continuum degrees of freedom of the waveguides gives an effective decay rate for each mode, $\kappa_n$, which increases monotonically as $\kappa_n \sim n^{0.3}$ (\Fig{subfig:KappaXrXl1Em3}). In the Supplementary Material, we show that for $\chi_s=0$ the resulting series \Eq{eqn:Multimode Purcell Rate} diverges \cite{Supplementary}, as pointed out in previous studies \cite{Houck_Controlling_2008, Filipp_Multimode_2011}. For {\it any} nonzero $\chi_s$, individual terms in the sum~(\ref{eqn:Multimode Purcell Rate}) display a universal power law $\sim n^{-2.7}$ (\Fig{subfig:LogLogSpEmRateXrXl1Em3}), which guarantees convergence \footnote{The power law dependence of $\kappa_n$ and $g_n$, though universal with respect to $\chi_s$, are specific to the chosen circuit topology.}.

\begin{figure}[t!]
\subfloat[\label{subfig:SpEmRateXrXl1Em3Xj5Em2Xg1Em3}]{%
\includegraphics[scale=0.31]{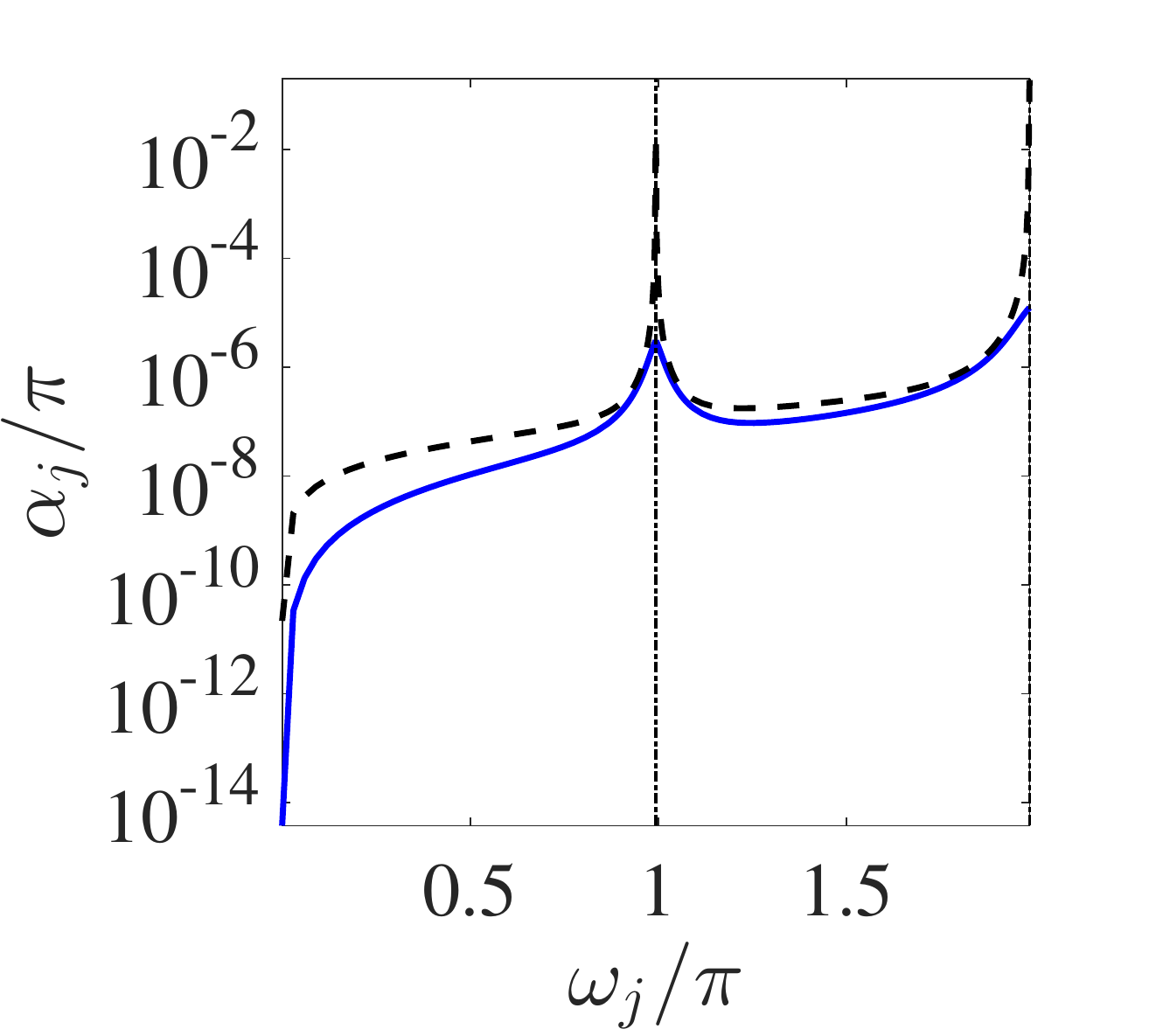}%
}
\hfill
\subfloat[\label{subfig:SpEmRateXrXl1Em3Xj5Em2Xg1Em1}]{%
\includegraphics[scale=0.31]{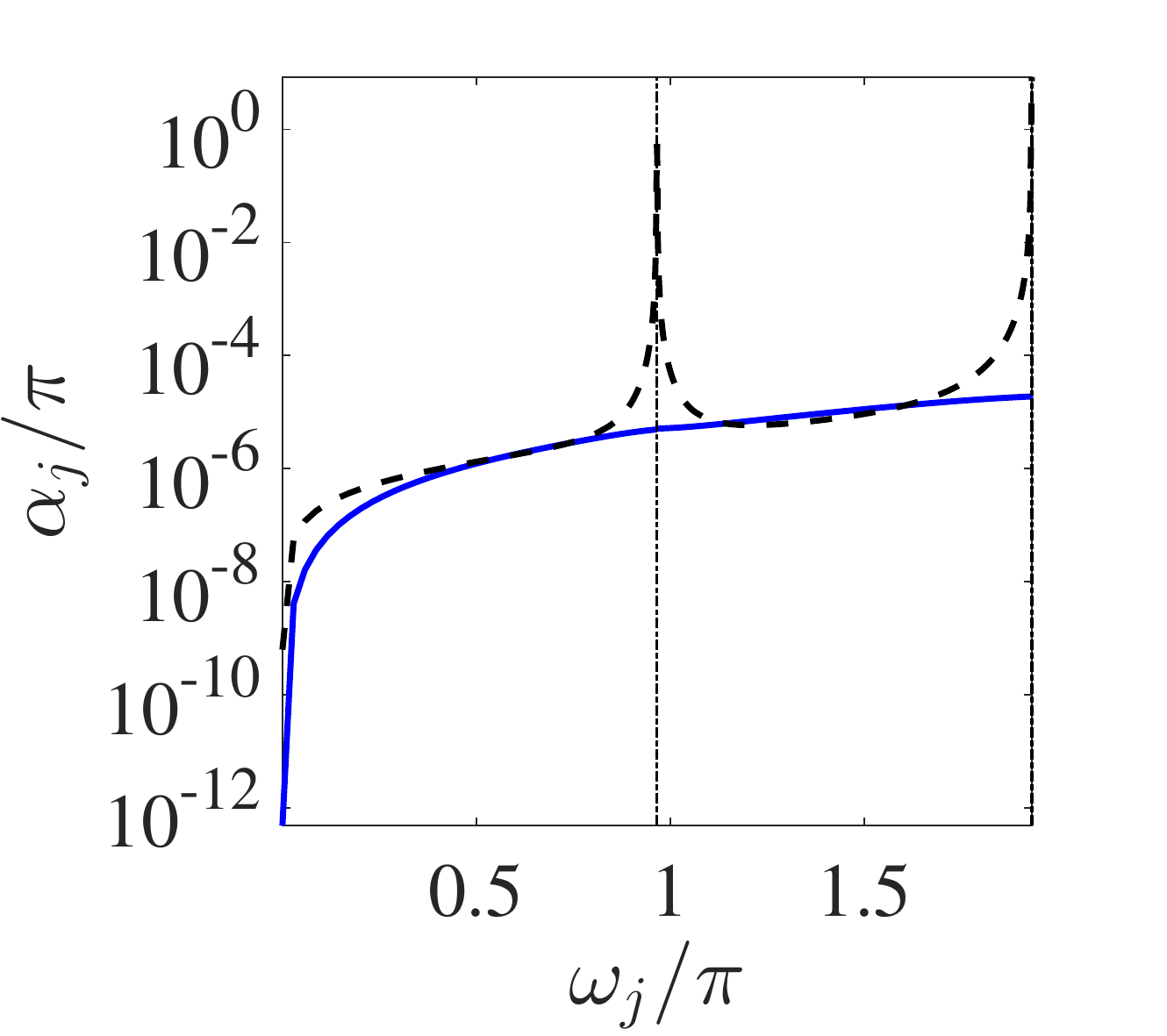}%
}\hfill
\subfloat[\label{subfig:LmbShftXrXl1Em3Xj5Em2Xg1Em3}]{%
\includegraphics[scale=0.31]{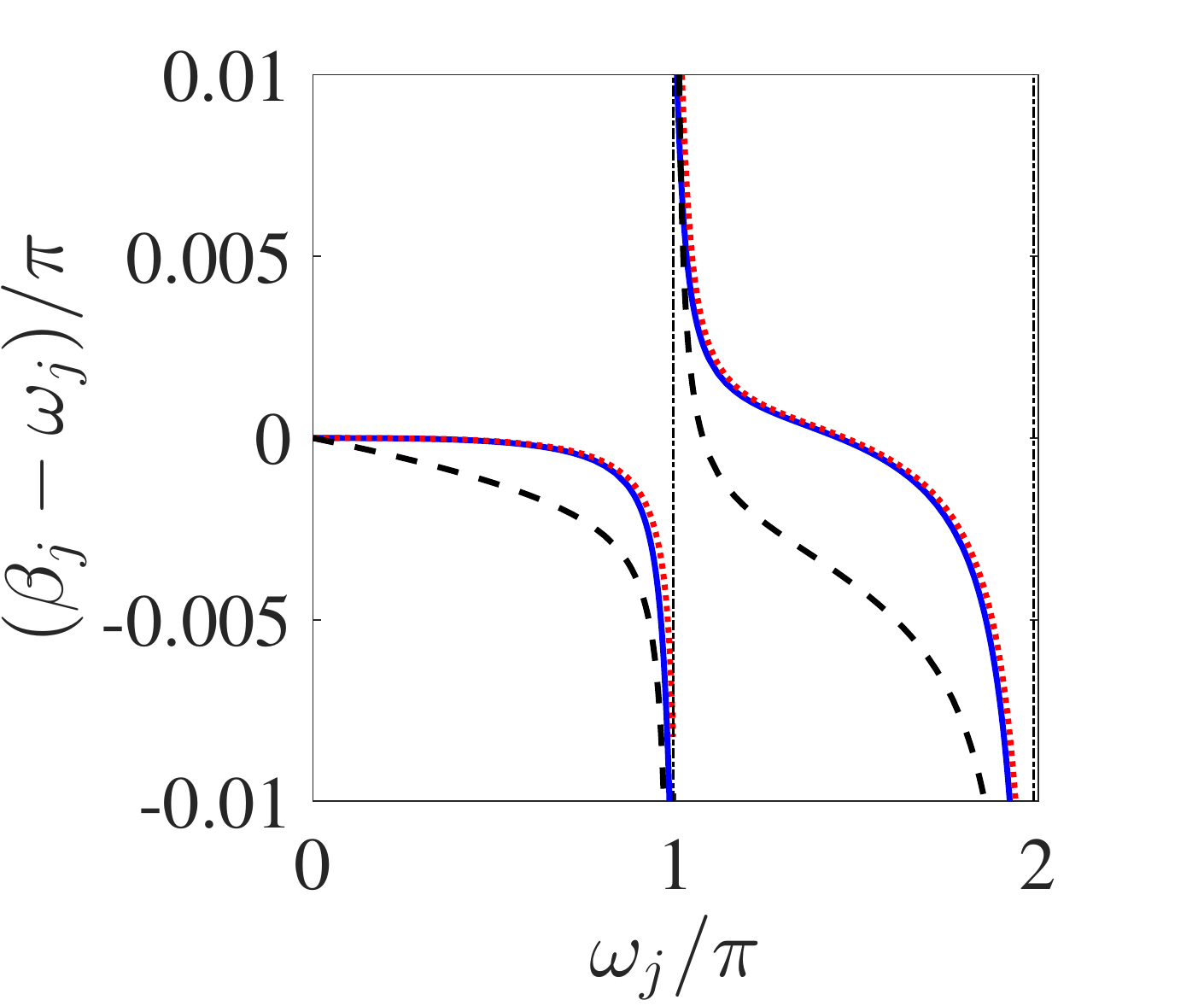}%
}
\hfill
\subfloat[\label{subfig:LmbShftXrXl1Em3Xj5Em2Xg1Em1}]{%
\includegraphics[scale=0.31]{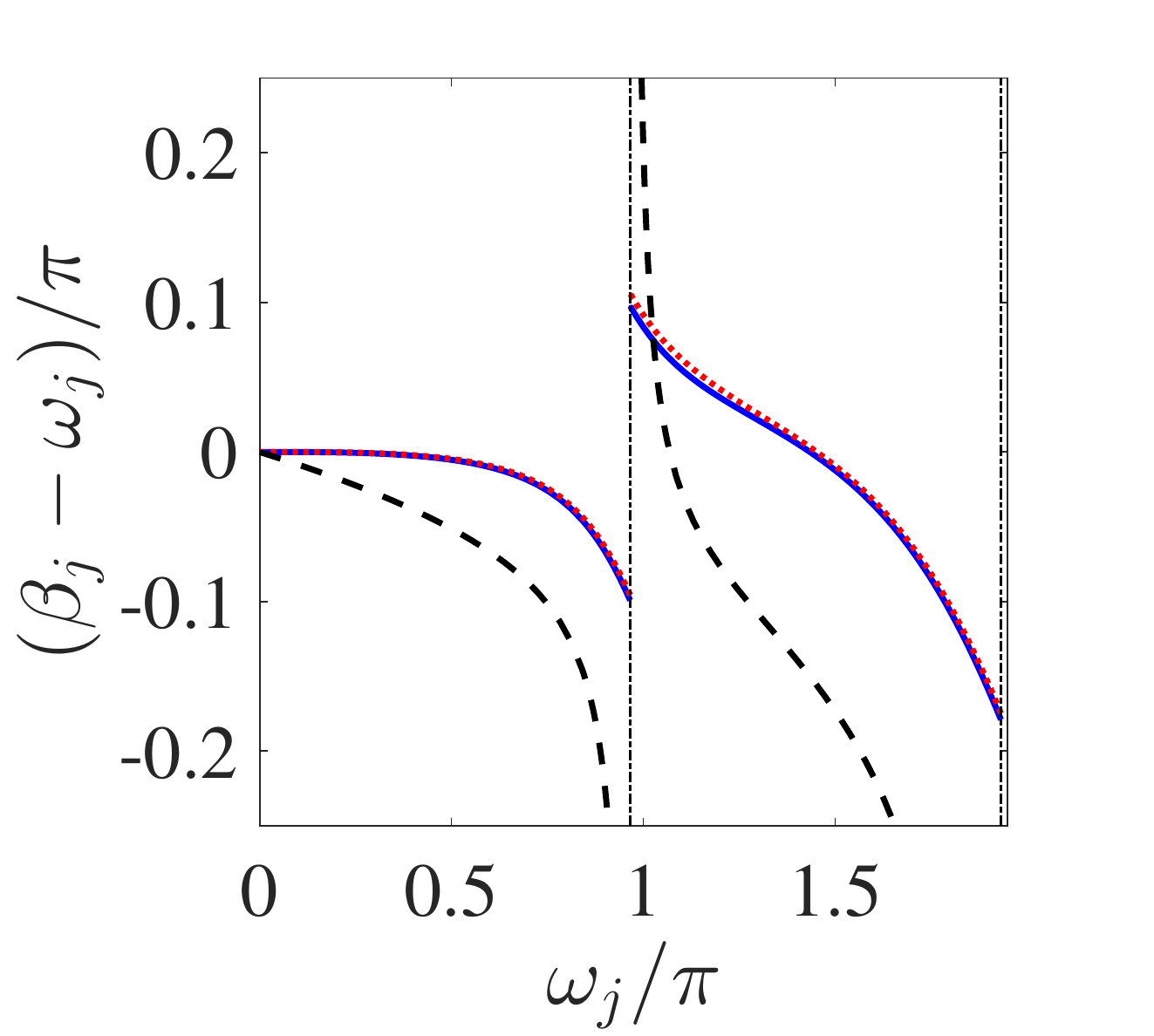}%
}
\caption{(Color online) Comparison of a,b) spontaneous decay rate between the linear theory (blue solid) and the dispersive limit result $\gamma_P$ (black dashed) as a function of $\omega_j$. c,d) Lamb shift between the linear theory (blue solid), leading order perturbation (red dotted) and the dispersive limit result $\Delta_L$ (black dashed). a,c) $\chi_g=0.001$ and b,d) $\chi_g=0.1$. Both values of $\chi_g$ are in strong coupling regime, i.e. $g_1/\alpha_j\gg 1$. However, $\chi_g=0.1$ ($g_1/\nu_1=0.1033$) reaches ultrastrong coupling \cite{Niemczyk_Circuit_2010}, where multimode effects are non-negligible. The nonlinearity is set as $\epsilon=0.1$, while other parameters are $\chi_R=\chi_L=10^{-3}$ and $\chi_j=0.05$. The vertical dash-dotted black line shows the position of the fundamental frequency of the resonator.} 
\label{Fig:SpEmRate&LambShift}
\end{figure}

\textit{Solution of the Heisenberg-Langevin equations.} Although we showed that the expression~(\ref{eqn:Multimode Purcell Rate}) for the Purcell decay rate converges, it is only valid in the dispersive regime $g_n \ll \delta_n$. This estimate for the Purcell decay rate and the Lamb shift will deviate substantially from the exact result for a range of order $g_n$ around each cavity resonance, diverging as the qubit frequency approaches the resonance (see Fig.~\ref{Fig:SpEmRate&LambShift}). This fictitious divergence can in principle be cured by solving the full multimode Master equation. Even if computational challenges relating to the long-time dynamics in such a large Hilbert space can be addressed, the resulting rate would still be subject to the TLA, RWA, Born and Markov approximations, casting a priori an uncertainty on its reliability.

An improved analytic result that is uniformly valid in the transmon frequency, and is not limited by the aforementioned approximations can be found by solving \Eqs{eqn:TransDyn}{eqn:ResDyn} perturbatively in the transmon's weak nonlinearity. EM degrees of freedom can be integrated out by solving \Eq{eqn:ResDyn} exactly, plugging into \Eq{eqn:TransDyn} and tracing over the photonic Hilbert space. To lowest order in the transmon nonlinearity $\epsilon = (E_c/E_j)^{1/2}$, where $E_c$ and $E_j$ are the charging and Josephson energy, respectively, the effective equation for the qubit is \cite{Malekakhlagh_NonMarkovian_2016}
\begin{align}
\begin{split}
\hat{\ddot{X}}_j(t)+\omega_j^2\left[1-\gamma+i\mathcal{K}_1(0)\right]\hat{X}_j(t)\\
=- \omega_j^2 \int_0^{t}dt' \mathcal{K}_2(t-t') \hat{X}_j(t'),
\end{split}
\label{eqn:LinSEProblem}
\end{align}
where $\hat{X}_j (t) =  \Tr_{ph}\{\hat{\rho}_{ph}(0)\hat{\varphi}_j(t)\}/\phi_{\text{zpf}}$ is the reduced flux operator traced over the photonic degrees of freedom and $\phi_{\text{zpf}}\equiv(\sqrt{2}\epsilon)^{1/2}$ is the magnitude of the zero-point phase fluctuations. This delay equation features the memory kernels  $\mathcal{K}_n(\tau)\equiv\gamma\chi_s\int_{-\infty}^{+\infty} \frac{d\omega}{2\pi} \, \omega^n \, G (x_0,x_0,\omega)e^{-i\omega\tau}$, where $G(x,x',\omega)$ is the classical EM Green's function defined by $\left[\partial_{x}^2-\chi(x,x_0)\partial_{t}^2\right] G(x,x',\omega) e^{-i \omega t} = e^{-i \omega t} \delta(x-x')$ implying that $G (x,x',\omega)$ is the amplitude of the flux field created at $x$ by a transmon oscillating with a frequency $\omega$ at $x'$ \cite{Malekakhlagh_NonMarkovian_2016}. The term on the right hand side of \ref{eqn:LinSEProblem} is therefore proportional to the fluctuating current driving the qubit at time $t$, that was excited by itself at an earlier time $t'$. This Green's function correctly encodes the modification of the capacitance per length. Equation~(\ref{eqn:LinSEProblem}) can be solved exactly in the Laplace domain
\begin{equation}
\hat{\tilde{X}}_j(s)=\frac{s\hat{X}_j(0)+ \hat{\dot{X}}_j(0)}{D_j(s)},
\label{eqn:Sol of X_j(s)}
\end{equation}
where $\tilde{h}(s)\equiv\int_{0}^{\infty} dt \, h(t) \, e^{-st}$, with $D_j(s)$ defined as \cite{Malekakhlagh_NonMarkovian_2016}
\begin{equation}
D_j(s)\equiv s^2+\omega_j^2\left[1-\gamma+i\mathcal{K}_1(0)+\tilde{\mathcal{K}}_2(s)\right].
\label{eqn:Def of D(s)}
\end{equation}
We express the characteristic function $D_j(s)$ in meromorphic form 
\begin{equation}
D_j(s)=(s-p_j)(s-p_j^*)\prod\limits_{m}\frac{(s-p_m)(s-p_m^*)}{(s-z_m)(s-z_m^*)}.
\label{eqn:Formal Rep of D(s)}
\end{equation}
The poles of $1/D_j(s)$ are the hybridized qubit-like and resonator-like complex-valued excitation frequencies, $p_j\equiv -\alpha_j-i\beta_j$ and $p_n\equiv -\alpha_n-i\beta_n$, respectively, of the qubit-resonator system, while its zeroes $z_n \equiv -i\omega_n=-\kappa_n-i\nu_n$ correspond to bare non-Hermitian \cite{Supplementary} cavity resonances. The real part of the qubit-like pole, $\alpha_j$, is the Purcell loss rate, while $\beta_j - \omega_j$ is the Lamb shift, as shown in \Fig{Fig:poles_sf_labeled}. In the Supplementary Material, we show that  $D_j(s)$ is convergent, and hence so are all hybridized frequencies, for any nonzero $\chi_s$.

The $A^2$-term kept in our calculation to enforce gauge invariance plays the role of the ``counterterm" discussed by Caldeira and Leggett to cancel infinite frequency renormalization \cite{Caldeira_Quantum_1983, Leggett_Quantum_1984}. This problem has also been discussed in the context of the quantum theory of laser radiation \cite{Schwabl_Quantum_1964}.

\textit{Perturbative corrections.} The transmon nonlinearity neglected in Eq.~(\ref{eqn:LinSEProblem}) can be reintroduced as a weak perturbation. The leading order correction to the hybridized resonances amounts to self- and cross-Kerr interactions \cite{Nigg_BlackBox_2012, Bourassa_Josephson_2012}. Using multi-scale perturbation theory \cite{Bender_Advanced_1999, Malekakhlagh_NonMarkovian_2016}, the correction to the transmon qubit-like resonance $\beta_j$ is given by
\begin{equation}
\hat{\beta}_j=\beta_j-\frac{\sqrt{2}\epsilon}{4}\omega_j\left[u_j^4\hat{\mathcal{H}}_j(0)+\sum\limits_n 2u_j^2u_n^2\hat{\mathcal{H}}_n(0)\right]
\label{eqn:PertCor}
\end{equation}
where the coefficients $u_{j,n}$ define the transformation from the hybridized to the unhybridized modes and $\hat{\mathcal{H}}_{j,n}(0)$ are the free Hamiltonians of the transmon and mode $n$, respectively. For $\chi_g\to0$, we find $u_j\to1$, $u_n=0$ and $\beta_j\to\omega_j$ such that we recover the frequency correction of free quantum Duffing oscillator $\hat{\bar{\omega}}_j=\omega_j[1-\frac{\sqrt{2}\epsilon}{4}\hat{\mathcal{H}}_j(0)]$ \cite{Bender_Multiple_1996}. We note three features of this result. Firstly, the correction is an operator and that expresses the fact that transmon levels are anharmonic. The anharmonicity can be calculated from the expectation value of a corrected quadrature operator \cite{Supplementary}. Secondly, by virtue of the lowest order result being convergent without a cutoff, the perturbative corrections are also convergent in the number of modes included. Finally, this result is not limited by the qubit-resonator coupling strength or the openness of the cavity. The final result is finite for all qubit frequencies, as opposed to the dispersive-limit result. The correction to the Purcell decay is higher order and forms the subject of future work.

We compared the spontaneous decay from the linear theory (blue solid) to the dispersive limit estimate $\gamma_P$ in Eq.~(\ref{eqn:Multimode Purcell Rate}) (black dashed) as the transmon frequency is tuned across the fundamental mode in Figs.~\ref{subfig:SpEmRateXrXl1Em3Xj5Em2Xg1Em3}-\ref{subfig:SpEmRateXrXl1Em3Xj5Em2Xg1Em1}. 
First, the spontaneous decay is asymmetric, since there are (in)finitely many modes with frequency (larger) smaller than $\omega_j$. This feature is captured by both theories. Second, the spontaneous decay is enhanced as the qubit frequency approaches the fundamental resonator frequency. However, the dispersive limit estimate is perturbative in $g_n/\delta_n$ and hence yields a divergent result (fake kink) on resonance regardless of coupling constant, contrary to our result~\ref{eqn:Formal Rep of D(s)} which predicts a finite value even at ultrastrong coupling (Fig.~\ref{subfig:SpEmRateXrXl1Em3Xj5Em2Xg1Em1} and caption).

In Figs.~\ref{subfig:LmbShftXrXl1Em3Xj5Em2Xg1Em3}-\ref{subfig:LmbShftXrXl1Em3Xj5Em2Xg1Em1} we compare the Lamb shift from the linear theory (blue solid) and the leading order perturbation theory (red dotted) to the dispersive multimode estimate (black dashed) $\sum_n g_n^2/\delta_n$ \cite{Boissonneault_Dispersive_2009}. Below the fundamental mode, the Lamb shift is negative due to the collective influence of all higher modes that redshifts the qubit frequency. Above the fundamental mode, there appears a competition between the hybridization with the fundamental mode and all higher modes. Close enough to the fundamental mode, the Lamb shift is positive until it changes sign, as predicted by all three curves.
 
\textit{Conclusion.} We have presented a framework to calculate the spontaneous decay and the Lamb shift of a transmon qubit, convergent in the number of resonator modes without the need for rotating-wave, two-level, Born or Markov approximations, or a high frequency cutoff. This is achieved by an \textit{ab initio} treatment of the quantum circuit equations of motion containing the $A^2$-term to enforce gauge invariance. Therefore, the modes of the resonator are modified such that the light-matter coupling is suppressed at high frequencies. Formulating the cavity resonances in terms of non-Hermitian modes provides access to the spontaneous decay, the Lamb shift, and any other QED observables in a unified way.

\textit{Acknowledgements.} We acknowledge helpful discussions with Zlatko Minev and S. M. Girvin. This work was supported by the US Department of Energy, Office of Basic Energy Sciences, Division of Materials Sciences and Engineering, under Award No. DE-SC0016011.

\textit{Note.} While finishing this manuscript we became aware of Ref.~\onlinecite{Gely_Convergence_2017}, which arrives at a similar conclusion for the Lamb shift in the dispersive regime through a different approach.  
\clearpage
\begin{center}
\textbf{\large Supplementary Material: Cutoff-free Circuit Quantum Electrodynamics}
\end{center}
\section{Heisenberg Equations Of Motion}
\label{Sec:QU EOM}
In this section, we present the Heisenberg equations of motion in terms of flux variables \cite{Bishop_Circuit_2010, Devoret_Quantum_2014}. These equations were derived before by the authors \cite{Malekakhlagh_NonMarkovian_2016} (see App.~A), but the main steps are summarized below for clarity. The flux variable is defined at any node $n$ in terms of the voltage at that node with respect to a fixed ground node
\begin{align}
\Phi_n(t)\equiv \int_{0}^{t}dt' V_n(t').
\label{Eq:Def of Phi_n(t)}
\end{align}
The classical Lagrangian is the sum of the Lagrangians for the Josephson junction, resonator, right and left waveguides, capacitive coupling between the resonator and the waveguides and the transmon-resonator capacitive coupling, respectively (let $U_j(\Phi_j)$ be the nonlinear Josephson potential):  
\begin{align}
\begin{split}
\mathcal{L}&=\underbrace{\frac{1}{2}C_j\dot{\Phi}_{j}(t)^2-U_j(\Phi_j(t))}_{\mathcal{L}_{j}} \\
&+\underbrace{\int_{0^+}^{L^-}\,dx\left[\frac{1}{2} c(\partial_{t}\Phi)^2-\frac{1}{2l}(\partial_{t}\Phi)^2\right]}_{\mathcal{L}_{\text{Res}}}\\
&+\underbrace{\int_{L^+}^{\infty}\,dx\left[\frac{1}{2} c(\partial_{t}\Phi_R)^2-\frac{1}{2l}(\partial_{x}\Phi_R)^2\right]}_{\mathcal{L}_{\text{RW}}}\\
&+\underbrace{\int_{-\infty}^{0^-}\,dx\left[\frac{1}{2} c(\partial_{t}\Phi_L)^2-\frac{1}{2l}(\partial_{x}\Phi_L)^2\right]}_{\mathcal{L}_{\text{LW}}}\\
&+\underbrace{\frac{1}{2}C_L\left[\dot{\Phi}_L(0^-,t)-\dot{\Phi}(0^+,t)\right]^2}_{\mathcal{L}_{C_L}}\\
&+\underbrace{\frac{1}{2}C_R\left[\dot{\Phi}_R(L^+,t)-\dot{\Phi}(L^-,t)\right]^2}_{\mathcal{L}_{C_R}}\\
&+\underbrace{\frac{1}{2}C_g\left[\dot{\Phi}_j(t)-\dot{\Phi}(x_0,t)\right]^2}_{\mathcal{L}_{C_g}},
\end{split}
\label{Eq:Full Lagrangian}
\end{align}
From Eq.~(\ref{Eq:Full Lagrangian}) one can derive, via a Legendre transformation followed by quantization \cite{Devoret_Quantum_1995, Clerk_Introduction_2010, Devoret_Quantum_2014}, the Hamiltonian operator associated with the quantum circuit. The quantum Hamiltonian for $C_{R,L}\to 0$ is in Ref. \onlinecite{Malekakhlagh_Origin_2016}. $C_{R,L} \neq 0$ leave equations of motion unchanged, but change boundary conditions (BCs) at $x=0,L$. Importantly, Heisenberg equations of motion for the quantum flux operators $\hat{\Phi}_j$, $\hat{\Phi}(x,t)$ and $\hat{\Phi}_{R,L}(x,t)$ turn out to be formally identical to Euler-Lagrange equations for~(\ref{Eq:Full Lagrangian}) with classical fields promoted to operators.

To express the Heisenberg equations of motion in a compact way, we introduce the following notations. $\Phi_0\equiv \frac{h}{2e}$ is the superconducting flux quantum and $E_j$ is the Josephson energy. $C_s \equiv C_gC_j/(C_g+C_j)$ is the series capacitance of $C_j$ and $C_g$ and  $\gamma\equiv C_g/(C_g+C_j)$. There is a modified capacitance per unit length in the resonator due to the coupling to the transmon qubit at position $x_0$:
\begin{equation}
  c(x,x_0)\equiv c+C_s\delta(x-x_0).
\end{equation}
$c$ and $l$ are the capacitance and inductance per unit length in the resonator and the waveguides.
 
We pass to unitless coordinates and operators ($v_p\equiv 1/\sqrt{lc}$)
\begin{align}
\begin{split}
&x\rightarrow\frac{x}{L},\quad
t\rightarrow\frac{t}{\frac{L}{v_{p}}},\quad
\omega\rightarrow\frac{\omega}{v_p}L,\\
&\hat{\varphi}\equiv 2\pi \frac{\hat{\Phi}}{\Phi_0}, \quad \hat{n}\equiv\frac{\hat{Q}}{2e}
\end{split}
\label{Eq:unitless vars}
\end{align}
The newly introduced operators $\hat{\varphi}$ and $\hat{n}$ represent phase and number and are canonically conjugate: $[\hat{\varphi}_j,\hat{n}_j]=i$ and $[\hat{\varphi}(x,t),\hat{n}(x',t')]=i\delta(x-x')\delta(t-t')$. Below we use unitless capacitances $\chi_i\equiv C_i/(cL)$, $i=R,L,j,g,s$, and the unitless capacitance per unit length becomes
\begin{align}
\chi(x,x_0)\equiv 1+\chi_s \delta(x-x_0).
\label{Eq:Def of chi(x,x0)}
\end{align}

In terms of the quantities introduced, the Heisenberg equations of motion for the superconducting phase operators are:
\begin{subequations}
\begin{align}
\hat{\ddot{\varphi}}_j(t)+(1-\gamma)\omega_j^2\sin{[\hat{\varphi}_j(t)]}=\gamma \partial_{t}^2\hat{\varphi}(x_0,t),
\label{Eq:Transmon Dyn}
\end{align}
\begin{align}
\begin{split}
\left[\partial_{x}^2-\chi(x,x_0)\partial_{t}^2\right]\hat{\varphi}(x,t)=\chi_s\omega_j^2 \sin{[\varphi_j(t)]}\delta(x-x_0),
\end{split}
\label{Eq:Res Dyn}
\end{align}
\begin{align}
\partial_{x}^2\hat{\varphi}_{R,L}(x,t)-\partial_{t}^2\hat{\varphi}_{R,L}(x,t)=0,
\label{Eq:Side Res Dyn}
\end{align}
\end{subequations}
with boundary conditions
\begin{subequations}
\begin{align}
\begin{split}
-\left.\partial_{x}\hat{\varphi}\right|_{x=1^-}&=-\left.\partial_{x}\hat{\varphi}_R\right|_{x=1^+}\\
&=\chi_R\partial_{t}^2\left[\hat{\varphi}(1^-,t)-\hat{\varphi}_R(1^+,t)\right],
\end{split}
\label{Eq:BC-Conservation of current at 1}
\end{align}
\begin{align}
\begin{split}
-\left.\partial_{x}\hat{\varphi}\right|_{x=0^+}&=-\left.\partial_{x}\hat{\varphi}_L\right|_{x=0^-}\\
&=\chi_L\partial_{t}^2\left[\hat{\varphi}_L(0^-,t)-\hat{\varphi}(0^+,t)\right],
\label{Eq:BC-Conservation of current at 0}
\end{split}
\end{align}
\begin{align}
\hat{\varphi}(x=x_0^-,t)=\hat{\varphi}(x=x_0^+,t),
\label{Eq:BC-Continuity of phase at x0}
\end{align}
\begin{align}
\begin{split}
\left. \partial_{x} \hat{\varphi}\right|_{x=x_0^+} &- \left.\partial_{x} \hat{\varphi} \right|_{x=x_0^-}
-\chi_s\partial_{t}^2 \hat{\varphi}(x_0,t)\\
&=\chi_s \omega_j^2 \sin{[\varphi_j(t)]}.
\end{split}
\label{Eq:BC-conservation of current at x0}
\end{align}
\end{subequations}

In Eqs.~(\ref{Eq:Transmon Dyn}) and (\ref{Eq:Res Dyn}), the oscillation frequency is unitless $\omega_j^2 = 8\mathcal{E}_c\mathcal{E}_j$, in terms of unitless Josephson and charging energies
\begin{align}
\mathcal{E}_{j,c}\equiv \sqrt{lc} L\frac{E_{j,c}}{\hbar}, \; E_c\equiv \frac{e^2}{2C_j}.
\end{align}
Equations~(\ref{Eq:Transmon Dyn}-\ref{Eq:Res Dyn}) are Eqs.~($2$-$3$) in the main text.

\section{Spectral Representation of the Green's function}
In this section we introduce a spectral representation of the Green's function. The Green's function enters the effective Heisenberg equation of motion for the superconducting phase of the transmon qubit (see Ref.~\onlinecite{Malekakhlagh_NonMarkovian_2016} for a complete derivation). The resonator Green's function appears if one follows this aim in Eqs.~(\ref{Eq:Transmon Dyn}) and~(\ref{Eq:Res Dyn}): one has to solve for $\hat{\varphi}(x,t)$, which is driven by the qubit in Eq.~(\ref{Eq:Res Dyn}), and substitute into~(\ref{Eq:Transmon Dyn}). The resonator Green's function is defined as the response of the resonator fields, described by the left hand sides of Eqs.~(\ref{Eq:Res Dyn}-\ref{Eq:Side Res Dyn}), to a $\delta$-function source in space-time
\begin{align}
\begin{split}
\left[\partial_x^2 -\chi(x,x_0)\partial_t^2\right]&G(x,t|x_0,t_0)=\delta(x-x_0)\delta(t-t_0),
\end{split}
\label{Eq:Def of G(x,t|x0,t0)}
\end{align}
obeying BCs~(\ref{Eq:BC-Conservation of current at 1}-\ref{Eq:BC-Continuity of phase at x0}) with $\hat{\varphi}(x,t)$ replaced by $G(x,t|x_0,t_0)$. Introducing Fourier transforms
\begin{subequations}
\begin{align}
&\tilde{G}(x,x_0,\omega)=\int_{-\infty}^{\infty}dt G(x,t|x_0,t_0) e^{+i\omega(t-t_0)}, \\
&G(x,t|x_0,t_0)=\int_{-\infty}^{\infty}\frac{d\omega}{2\pi} \tilde{G}(x,x_0,\omega)e^{-i\omega(t-t_0)},
\end{align}
\end{subequations}
Equation~(\ref{Eq:Def of G(x,t|x0,t0)}) becomes a Helmholtz equation 
\begin{align}
\begin{split}
\left[\partial_x^2 +\omega^2\chi(x,x_0) \right]\tilde{G}(x,x_0,\omega)=\delta(x-x_0).
\end{split}
\label{Eq:Helmholtz Eq for G(x,x0,W)}
\end{align}
while the BCs~(\ref{Eq:BC-Conservation of current at 1}-\ref{Eq:BC-Continuity of phase at x0}) are transformed by replacing $\partial_t \to -i\omega $ to
\begin{subequations}
\begin{align}
\begin{split}
\left.\partial_x\tilde{G}\right|_{x=1^-}&=\left.\partial_x \tilde{G}\right|_{x=1^+}\\
&=\chi_R \omega^2 \left(\left.\tilde{G}\right|_{x=1^-}-\left.\tilde{G}\right|_{x=1^+} \right), 
\end{split}
\label{Eq:Cont of dxG(x,x0,W) at 1}
\end{align}
\begin{align}
\begin{split}
\left.\partial_x \tilde{G}\right|_{x=0^-}&= \left.\partial_x \tilde{G}\right|_{x=0^+}\\
&=\chi_L \omega^2 \left(\left.\tilde{G}\right|_{x=0^-}-\left.\tilde{G}\right|_{x=0^+} \right). 
\end{split}
\label{Eq:Cont of dxG(x,x0,W) at 0}
\end{align}
\begin{align}
&\left.\tilde{G}\right|_{x=x_0^+}=\left.\tilde{G}\right|_{x=x_0^-}, 
\label{Eq:Cont of G(x,x0,W) at x0}\\
&\left.\partial_x \tilde{G} \right|_{x=x_0^+}-\left.\partial_x \tilde{G}\right|_{x=x_0^-}+\chi_s\omega^2\left.\tilde{G}\right|_{x=x_0}=1,
\label{Eq:Cont of dxG(x,x0,W) at x0}
\end{align}
\end{subequations}
Lastly, outgoing BCs at infinity model the baths:
\begin{align}
\left. \partial_x \tilde{G}(x,x_0,\omega)\right|_{x\to \pm\infty}=\pm i\omega \tilde{G}(x\to\pm\infty,x_0,\omega).
\label{Eq:Outgoing BC for G(x,x0,W))}
\end{align}
Excitations leaving the resonator never reflect back towards it.

\subsection{Spectral representation of Green's function for $\chi_{R,L} = 0$}
\label{SubApp:Spec Rep of G-closed}
Setting $\chi_R=\chi_L=0$ (amounting to a closed resonator) imposes Neumann BC $\partial_x \tilde{G}|_{x=0,1}=0$ and the problem for $\tilde{G}$ is Hermitian. $\tilde{G}$ can be expanded in terms of a discrete set of normal modes satisfying
\begin{subequations}
\begin{align}
&\partial_x^2\tilde{\varphi}_n(x)+\chi(x,x_0)\omega_n^2\tilde{\varphi}_n(x)=0,
\label{Eq:Helmholtz Eq for Phi_n(x)}\\
&\left.\partial_x \tilde{\varphi}_n(x)\right|_{x=0,1}=0.
\label{Eq:Closed Neumann BC}
\end{align}
\end{subequations}
An important feature of the modes is that their derivative is discontinuous
\begin{align}
-\left.\partial_x \tilde{\varphi}_n(x)\right|_{x_0^-}^{x_0^+}=\chi_s\omega_n^2 \tilde{\varphi}_n(x_0),
\label{Eq:Current-Conservation} 
\end{align}
Physically, this is the continuity equation at $x_0$, or current conservation. The mode amplitude at $x_0$ is suppressed. These observations lead us to name this set of resonator eigenmodes the {\it current-conserving (CC) basis}. 

The CC basis eigenfrequencies obey a transcendental equation
\begin{align}
\sin{(\omega_n)}+\chi_s\omega_n\cos{(\omega_n x_0)}\cos{\left[\omega_n (1-x_0)\right]}=0,
\label{Eq:Hermitian Eigenfrequencies}
\end{align}
while the eigenfunctions read
\begin{align}
\tilde{\varphi}_n(x)\propto
\begin{cases}
\cos{\left[\omega_n (1-x_0)\right]}\cos{(\omega_n x)},&0<x<x_0\\
\cos{(\omega_n x_0)}\cos{\left[\omega_n (1-x)\right]},&x_0<x<1
\end{cases}
\label{Eq:Sol of Phi_n(x)-closed}
\end{align}
and the basis is orthonormal over $[0,1]$: 
\begin{align}
\int_{0}^{1}dx\chi(x,x_0)\tilde{\varphi}_m(x)\tilde{\varphi}_n(x)=\delta_{mn}.
\label{Eq:Closed Orthogonality Condition}
\end{align}
Equation~(\ref{Eq:Hermitian Eigenfrequencies}) can be solved numerically or asymptotically as $n\to\infty$, as we do in Sec.~\ref{App:Asymptotic of Eig}. 

The spectral representation of $\tilde{G}(x,x',\omega)$ \cite{Morse_Methods_1953, Economou_Green_1984, Hassani_Mathematical_2013} is
\begin{align}
\tilde{G}(x,x',\omega)=\sum\limits_{n\in \mathbb{N}}\frac{\tilde{\varphi}_n(x)\tilde{\varphi}_n(x')}{\omega^2-\omega_n^2}=\sum\limits_{n\in \mathbb{Z}\atop n\neq 0}\frac{1}{2\omega}\frac{\tilde{\varphi}_n(x)\tilde{\varphi}_n(x')}{\omega-\omega_n},
\label{Eq:Spectral rep of G-closed}
\end{align}
since $\omega_{-n}=-\omega_{n}$ and $\tilde{\varphi}_{-n}(x)=\tilde{\varphi}_{n}(x)$.
\subsection{Spectral representation of Green's function for $\chi_{R,L} \neq 0$}
\label{SubApp:Spec Rep of G-open} 
If the resonator is open, $\chi_{L,R} \neq 0$, we resort to a spectral representation in terms of a discrete set of non-Hermitian modes \cite{Malekakhlagh_NonMarkovian_2016} that carry constant flux away from the resonator, Constant Flux (CF) modes \cite{Tureci_SelfConsistent_2006,*Tureci_Strong_2008}. CF modes satisfy the homogeneous wave equation~
\begin{align}
\partial_x^2\tilde{\varphi}_n(x,\omega)+\chi(x,x_0)\omega_n^2(\omega)\tilde{\varphi}_n(x,\omega)=0,
\label{Eq:Helmholtz Eq for Phi_n(x)}
\end{align}
with BCs~(\ref{Eq:Cont of dxG(x,x0,W) at 1})-(\ref{Eq:Cont of G(x,x0,W) at x0})~and~(\ref{Eq:Outgoing BC for G(x,x0,W))}). Both the modes $\tilde{\varphi}_n(x,\omega)$ and their frequencies $\omega_n(\omega)$ depend on the source frequency $\omega$. 

An outgoing plane wave solution for the left and right waveguides obeying~(\ref{Eq:Outgoing BC for G(x,x0,W))}), is
\begin{align}
\tilde{\varphi}_n(x,\omega)=
\begin{cases}
A_{n}^<e^{i\omega_n(\omega) x}+B_{n}^< e^{-i\omega_n(\omega) x},&0<x<x_0\\
A_{n}^>e^{i\omega_n(\omega) x}+B_{n}^> e^{-i\omega_n(\omega) x},&x_0<x<1\\
C_ne^{i\omega x},&x>1 \\
D_ne^{-i\omega x},&x<0 \\
\end{cases}
\label{Eq:General Ansantz for Phi_n(x)}
\end{align}
Applying BCs (\ref{Eq:Cont of G(x,x0,W) at x0}-\ref{Eq:Cont of dxG(x,x0,W) at 0}) leads to a transcendental equation analogous to the closed case which fixes the parametric dependence $\omega_n(\omega)$ \cite{Malekakhlagh_NonMarkovian_2016}.  

The CF modes satisfy now a biorthonormality \cite{Tureci_SelfConsistent_2006} condition
\begin{align}
\begin{split}
\int_{0}^{1}dx\chi(x,x_0)\bar{\tilde{\varphi}}_m^*(x,\omega)\tilde{\varphi}_n(x,\omega)=\delta_{mn},
\label{Eq:Open Ortho Cond-unsimp}
\end{split}
\end{align}
where $\{\bar{\tilde{\varphi}}_m(x,\omega)\}$ obey the Hermitian adjoint of~(\ref{Eq:Helmholtz Eq for Phi_n(x)}). $\tilde{\varphi}_n(x,\omega)$ and $\bar{\tilde{\varphi}}_n(x,\omega)$ are right and left eigenfunctions and obey $\bar{\tilde{\varphi}}_n(x,\omega)=\tilde{\varphi}_n^*(x,\omega)$.

The CF mode spectral representation of the Green's function of the open resonator is
\begin{align}
\tilde{G}(x,x',\omega)=\sum\limits_{n}\frac{\tilde{\varphi}_n(x,\omega)\bar{\tilde{\varphi}}_n^*(x',\omega)}{\omega^2-\omega_n^2(\omega)}.
\label{Eq:Spectral rep of G-Open}
\end{align}
There are two sets of poles of $\tilde{G}(x,x',\omega)$ in the complex plane. When the denominator of~(\ref{Eq:Spectral rep of G-Open}) vanishes, $\omega=\omega_n(\omega)$, which corresponds to quasi-bound eigenfrequencies that obey  
\begin{align}
\begin{split}
&\left[e^{2i\omega_n}-(1-2i\chi_L\omega_n)(1-2i\chi_R\omega_n)\right]\\
&+\frac{i}{2}\chi_s\omega_n[e^{2i\omega_n x_0}+(1-2i\chi_L\omega_n)]\\
&\times[e^{2i\omega_n (1-x_0)}+(1-2i\chi_R\omega_n)]=0.
\end{split}
\label{Eq:Generic NHEigfreq}
\end{align}
The solutions reside in the lower half of complex $\omega$-plane and come in symmetric pairs with respect to the $\Im\{\omega\}$ axis, i.e. if $\omega_{n}$  satisfies~(\ref{Eq:Generic NHEigfreq}), so does $-\omega_{n}^*$. Therefore the eigenfrequencies are
\begin{align}
\omega_n=\begin{cases}
-i\kappa_0, \quad & n=0\\
+\nu_n-i\kappa_n, \quad &n\in+\mathbb{N}\\
-\nu_n-i\kappa_n, \quad &n\in-\mathbb{N}
\end{cases}
\end{align}
where $\nu_n > 0$ and $\kappa_n >0$ are the oscillation frequency and decay rates of quasi-bound mode $n$, respectively. The dependence of $\kappa_n$ on mode number $n$ is plotted in Fig.~$2$ of the main letter. Note the existence of a pole at $\omega=0$, which comes from the $\omega$-dependence of CF states $\tilde{\varphi}_n(x,\omega)$ \cite{Tureci_SelfConsistent_2006,*Tureci_Strong_2008}. 
\section{Multimode Jaynes-Cummings Hamiltonian}
\label{App:MultiJC}
The classical Hamiltonian for the cQED system can be found from the circuit Lagrangian~(\ref{Eq:Full Lagrangian}) \cite{Malekakhlagh_Origin_2016}
\begin{align}
\begin{split}
\mathcal{H}_{sys}&=4\mathcal{E}_c n_j^2(t)-\mathcal{E}_j\cos{[\varphi_j(t)]}\\
&+\int_{0}^{1}dx \left\{\frac{n^2(x,t)}{2\chi(x,x_0)}+\frac{1}{2}\left[\partial_x \varphi(x,t)\right]^2\right\}\\
&+2\pi\gamma z n_j(t)\int_{0}^{1}dx\frac{n(x,t)}{\chi(x,x_0)}\delta(x-x_0),
\end{split}
\label{Eq:model-closed cQED H}
\end{align}
where $z\equiv Z/R_Q$ where $Z\equiv\sqrt{l/c}$ is the characteristic impedance of the resonator and $R_Q\equiv h/(2e)^2$ is the superconducting resistance quantum. The modification in capacitance per length originates from the system Lagrangian that contains the gauge-invariant qubit-resonator coupling $\chi_g[\dot{\varphi}_j(t)-\dot{\varphi}(x_0,t)]^2/2$. In contrast, a phenomenological product coupling $\chi_g\dot{\varphi}_j(t)\dot{\varphi}(x_0,t)$ would yield a $\mathcal{H}_{sys}$ with $\chi_s=0$ which results in bare resonator modes.

For the purpose of quantizing $\mathcal{H}_{sys}$, we find the spectrum of the resonator by solving the corresponding Helmholtz eigenvalue problem that has been discussed in Sec.~(\ref{SubApp:Spec Rep of G-closed}). We find

\begin{align}
\begin{split}
&\hat{\mathcal{H}}_{\text{sys}} \equiv \frac{\omega_j}{4}\left\{\hat{\mathcal{Y}}_j^2-\frac{\sqrt{2}}{\epsilon}\cos\left[(2\epsilon^2)^{1/4}\hat{\mathcal{X}}_j\right]\right\}\\
&+\sum\limits_{n}\left\{\frac{\nu_n}{4}\left[\hat{\mathcal{X}}_n^2+\hat{\mathcal{Y}}_n^2\right]+g_n\hat{\mathcal{Y}}_j\hat{\mathcal{Y}}_n\right\},
\end{split}
\label{Eq:2nd quantized closed H}
\end{align}
where have defined the canonically conjugate variables $\hat{\mathcal{X}}_l\equiv(\hat{a}_l+\hat{a}_l^{\dag})$ and $\hat{\mathcal{Y}}_l\equiv -i(\hat{a}_l-\hat{a}_l^{\dag})$, where $\hat{a}_{l}$ represent the boson annihilation operator of sector $l\equiv j,c$. Moreover, $\omega_j\equiv \sqrt{8\mathcal{E}_j\mathcal{E}_c}$ and $\epsilon\equiv\sqrt{\mathcal{E}_c/\mathcal{E}_j}$ is a measure for the strength of transmon nonlinearity. For $\epsilon=0$, we recover $\omega_j(\hat{\mathcal{X}}_j^2+\hat{\mathcal{Y}}_j^2)/4$, the Hamiltonian of a simple harmonic oscillator. In the transmon regime where $\epsilon\ll 1$, the leading contribution is $-\sqrt{2}\epsilon\omega_j\hat{\mathcal{X}}_j^4/48$. The coupling between qubit and the $n$th CC mode of the resonator is
\begin{align} 
g_n=\frac{1}{2}\gamma\sqrt{\chi_j}\sqrt{\omega_j\nu_n}\tilde{\varphi}_n(x_0).
\label{Eq:Light-Matter g_n} 
\end{align}

There are typically two approaches to diagonalize Eq.~(\ref{Eq:2nd quantized closed H}). In the first approach, assuming that the qubit nonlinearity is strong, one performs a two level reduction. Then, the multimode Rabi Hamiltonian can be derived from Eq.~(\ref{Eq:2nd quantized closed H}) by projecting the quadratures to Pauli sigma matrices, $\hat{\mathcal{X}}_j\rightarrow\hat{\sigma}^x$ and $\hat{\mathcal{Y}}_j\rightarrow\hat{\sigma}^y$, which yields
\begin{align}
\begin{split}
\hat{\mathcal{H}}_{\text{Rabi}}&=\frac{\omega_j}{2}\hat{\sigma}^z+\sum\limits_{n}\nu_n\hat{a}_n^{\dag}\hat{a}_n\\
&-\sum\limits_ng_n(\hat{a}_n-\hat{a}_n^{\dag})(\hat{\sigma}^--\hat{\sigma}^+).
\end{split}
\label{Eq:Rabi H}
\end{align}
In the rotating wave approximation, Eq.~(\ref{Eq:Rabi H}) transforms into the multimode Jaynes-Cummings Hamiltonian
\begin{align}
\begin{split}
\hat{\mathcal{H}}_{\text{JC}}&=\frac{\omega_j}{2}\hat{\sigma}^z+\sum\limits_{n}\nu_n\hat{a}_n^{\dag}\hat{a}_n+\sum\limits_ng_n(\hat{\sigma}^+\hat{a}_n+\hat{\sigma}^-\hat{a}_n^{\dag})
\end{split}
\label{Eq:JC H}
\end{align}
used in the main text. Analytic results can be found for the Purcell decay rate and the Lamb shift in the dispersive limit where $g_n\ll|\omega_j-\omega_n|$ \cite{Boissonneault_Dispersive_2009}. In a Lindblad calculation, resonator losses are included by a Bloch-Redfield approach through the Master equation for the reduced density matrix of the resonator and qubit degrees of freedom $\hat{\dot{\rho}} = -i [ \hat{\mathcal{H}}_\text{JC}, \hat{\rho} ] + \frac{\kappa_n}{2} \left( 2 \hat{a}_n \hat{\rho} \hat{a}_n^\dagger - \{ \hat{\rho}, \hat{a}_n^\dagger \hat{a}_n \} \right)$, where $\kappa_n$ can be replaced from the solutions to Eq.~(\ref{Eq:Generic NHEigfreq}). The second approach treats the nonlinearity as a weak perturbation and is explained in the next section.
\section{Weakly Nonlinear Transmon}
In this section we summarize the steps necessary to derive Eq.~(10) of the main text. The full development of multi scale perturbation theory is in Ref.~\onlinecite{Malekakhlagh_NonMarkovian_2016}. By keeping the lowest order nonlinearity (Kerr terms which are quartic in the transmon quadrature), the Hamiltonian can be rewritten in a new basis that diagonalizes the quadratic part
\begin{align}
\begin{split}
\hat{\mathcal{H}}_{\text{sys}} &\equiv \frac{\beta_j}{4}\left(\hat{\bar{\mathcal{X}}}_j^2+\hat{\bar{\mathcal{Y}}}_j^2\right)+\sum\limits_n\frac{\beta_n}{4}\left(\hat{\bar{\mathcal{X}}}_n^2+\hat{\bar{\mathcal{Y}}}_n^2\right)\\
&-\frac{\varepsilon\omega_j}{8}\left(u_j \hat{\bar{\mathcal{X}}}_j+\sum\limits_nu_n \hat{\bar{\mathcal{X}}}_n\right)^4,
\end{split}
\label{Eq:QuDuffQuHarm-H NormModes}
\end{align}
where $\varepsilon\equiv \sqrt{2}\epsilon/6$, $\beta_{j,n}$ are hybridized frequencies and $u_{j,n}$ are hybridization coefficients: $\hat{\mathcal{X}}_j=u_j\hat{\bar{\mathcal{X}}}_j+\sum\limits_n u_n\hat{\bar{\mathcal{X}}}_n$.

The Heisenberg equations of motion for quadratures become a set of quantum Duffing equations coupled via the quartic terms
\begin{align}
\begin{split}
\hat{\ddot{\bar{\mathcal{X}}}}_{l}(t)+\beta_{l}^2\left\{\hat{\bar{\mathcal{X}}}_{l}(t)-\varepsilon_{l} \left[u_j\hat{\bar{\mathcal{X}}}_j(t)+\sum\limits_n u_n\hat{\bar{\mathcal{X}}}_n(t)\right]^3\right\}=0,
\end{split}
\label{Eq:QuDuffQuHarm Osc}
\end{align}
where $\varepsilon_{l}\equiv\frac{\omega_j}{\beta_{l}}u_{l}\varepsilon$ for $l\equiv j,n$. Up to lowest order in the perturbation \cite{Malekakhlagh_NonMarkovian_2016}, we find an operator valued correction of the linear theory qubit-like frequency $\beta_j$: 
\begin{subequations}
\begin{align}
\hat{\beta}_{j}=\beta_{j}-\frac{\sqrt{2}\epsilon}{4}\omega_j\left[u_j^4\hat{\bar{\mathcal{H}}}_j(0)+\sum\limits_{n}2u_j^2u_n^2\hat{\bar{\mathcal{H}}}_n(0)\right],
\label{Eq:QuDuffQuHarm-bar(beta)_j}
\end{align}
and an analogous correction of the resonator like frequency $\beta_{n}$ as
\begin{align}
\begin{split}
\hat{\beta}_{n}=\beta_{n}-\frac{\sqrt{2}\epsilon}{4}\omega_j&\left[u_n^4\hat{\bar{\mathcal{H}}}_n(0)+2u_n^2u_j^2\hat{\bar{\mathcal{H}}}_j(0)\right.\\
&+\left.\sum\limits_{m\neq n}2u_n^2u_m^2\hat{\bar{\mathcal{H}}}_m(0)\right],
\end{split}
\label{Eq:QuDuffQuHarm-bar(beta)_n}
\end{align}
\end{subequations}
where $\hat{\bar{\mathcal{H}}}_l(0)\equiv\frac{1}{4}[\hat{\bar{\mathcal{X}}}_l^2(0)+\hat{\bar{\mathcal{Y}}}_l^2(0)]$ for $l=j,n$. In the main text, Eq.~($10$), the bar notation is dropped. The lowest order MSPT solution for the qubit quadrature becomes, in terms of renormalized frequencies $\hat{{\beta}}_{j,n}$, \cite{Malekakhlagh_NonMarkovian_2016}
\begin{align}
\begin{split}
\hat{\mathcal{X}}_j^{(0)}(t)&=u_j\frac{\hat{{a}}_j(0)e^{-i\hat{\beta}_j t}+e^{-i\hat{{\beta}}_j t}\hat{{a}}_j(0)}{2\cos\left(\frac{3\omega_j}{4}u_j^4\varepsilon t\right)}+H.c.\\
&+\sum\limits_n\left[u_n\frac{\hat{{a}}_n(0)e^{-i\hat{{\beta}}_n t}+e^{-i\hat{{\beta}}_n t}\hat{{a}}_n(0)}{2\cos\left(\frac{3\omega_j}{4}u_n^4\varepsilon t \right)}+H.c.\right].
\label{Eq:QuDuffQuHarm-X_j^(0)(t) MSPT Sol}
\end{split}
\end{align}
This equation takes into account corrections up to $\mathcal{O}(\varepsilon)$ in frequencies. To extract these corrections, we must evaluate the expectation value of Eq.~(\ref{Eq:QuDuffQuHarm-X_j^(0)(t) MSPT Sol}) with respect to the initial density matrix. We chose $\hat{\rho}=\ket{\Psi}_j\bra{\Psi}_j\otimes\ket{0}_{\text{ph}}\bra{0}_{\text{ph}}$ with $\ket{\Psi}_j=(\ket{0}_j+\ket{1}_j)/\sqrt{2}$. The correction to the transmon like frequency is obtained from the Fourier components of $\left<\hat{\mathcal{X}}_j(t)\right>$. This is the correction plotted in Fig.~$3$ of the main text.
\section{Asymptotic Behavior of Light-Matter coupling}
\label{App:Asymptotic of Eig}
In this section we find the asymptotic behavior of the eigenfrequencies $\omega_n$ and eigenmodes $\tilde{\varphi}_n(x)$ of the resonator discussed in the main text. This provides an analytical understanding of the high frequency suppression in the light-matter coupling $g_n$. 

To point out the origin of the suppression that arise from a nonzero $\chi_s$, let us consider the closed resonator ($\chi_{R,L}=0$) case. Consider the special case of $x_0=0^+$ first. This is of experimental interest in order to achieve the maximum coupling to all modes of a resonator. Then, the transcendental Eq.~(\ref{Eq:Hermitian Eigenfrequencies}) simplifies to
\begin{align}
\sin(\omega_n)+\chi_s\omega_n\cos(\omega_n)=0,
\label{Eq:HermEigFreq-x0=0}
\end{align}
which can be rewritten as 
\begin{align}
\tan(\omega_n)=-\chi_s\omega_n.
\label{Eq:HermEigFreq-x0=0-Simplified}
\end{align}
The large $\omega_n$ solution for $\chi_s\neq0$ is then obtained 
\begin{align}
\lim_{n\to\infty}\omega_n=n\pi-\frac{\pi}{2},
\end{align} 
which is independent of the value for $\chi_s$. This implies that the effect of a nonzero $\chi_s$ on $\omega_n$ is a total shift $\pi/2$ (half of the free spectral range) in comparison with the case $\chi_s=0$. Substituting $x_0=0^+$ in Eq.~(\ref{Eq:Sol of Phi_n(x)-closed}), the normalization factor $\mathcal{N}_n$ is found via Eq.~(\ref{Eq:Closed Orthogonality Condition}) as
\begin{align}
\int_0^{1}dx\cos^2[\omega_n(1-x)]+\chi_s\cos^2(\omega_n)=\frac{1}{\mathcal{N}_n^2},
\end{align}
which gives
\begin{align}
\mathcal{N}_n=\frac{\sqrt{2}}{\sqrt{1+\chi_s\cos^2(\omega_n)}}.
\label{Eq:Normalization_x0=0}
\end{align}
Therefore the eigenmode is found as
\begin{align}
\tilde{\varphi}_n(x_0=0^+)=\frac{\sqrt{2}\cos(\omega_n)}{\sqrt{1+\chi_s\cos^2(\omega_n)}}.
\label{Eq:HermEigMode@x0-x0=0}
\end{align}
Using the trigonometric identity 
\begin{align}
\cos^2(\omega_n)=\frac{1}{1+\tan^2(\omega_n)}
\end{align}
and Eq.~(\ref{Eq:HermEigFreq-x0=0-Simplified}) we can rewrite Eq.~(\ref{Eq:HermEigMode@x0-x0=0}) as
\begin{align}
\tilde{\varphi}_n(x_0=0^+)=\frac{\sqrt{2}}{\sqrt{1+\chi_s+\chi_s^2\omega_n^2}},
\label{Eq:HermEigMode@x0-x0=0-Simplified}
\end{align}
which now provides the algebraic dependence of $\tilde{\varphi}_n(x_0)$ on $\omega_n$. According to Eq.~(\ref{Eq:HermEigMode@x0-x0=0-Simplified}), for large enough $\omega_n$ ($\chi_s\omega_n \gg 1+\chi_s$), we find
\begin{align}
\tilde{\varphi}_n(x_0)\thicksim\frac{1}{\omega_n},
\label{Eq:AsymptOfPhi_n}
\end{align}
where the symbol $\thicksim$ represents asymptotic equivalence. This imposes a natural cut-off on the light matter coupling for $n\to\infty$, since
\begin{align}
g_n\propto \sqrt{\omega_n}\tilde{\varphi}_n(x_0)\thicksim \frac{1}{\sqrt{\omega_n}}.
\label{Eq:AsymptOfg_n}
\end{align}

Next, we would like to find the asymptotic behavior of $\omega_n$ and $\tilde{\varphi}_n(x_0)$ for a general $x_0$. In order to bring Eq.~(\ref{Eq:Hermitian Eigenfrequencies}) into a similar form to Eq.~(\ref{Eq:HermEigFreq-x0=0-Simplified}), we first replace $\sin(\omega_n)=\sin[\omega_nx_0+\omega_n(1-x_0)]$ and then divide by $\cos(\omega_n x_0)\cos[\omega_n (1-x_0)]$ to obtain
\begin{align}
\tan(\omega_n x_0)+\tan[\omega_n (1-x_0)]=-\chi_s\omega_n.
\label{Eq:HermEigFreq-Simplified}
\end{align}
Next, the normalization factor $\mathcal{N}_n$ is found from Eq.~(\ref{Eq:Closed Orthogonality Condition}) as
\begin{widetext}
\begin{align}
\mathcal{N}_n=\frac{\sqrt{2}}{\sqrt{x_0\cos^2[\omega_n(1-x_0)]+(1-x_0)\cos^2(\omega_n x_0)+\chi_s\cos^2[\omega_n(1-x_0)]\cos^2(\omega_nx_0)}},
\label{Eq:Normalization-Simplified}
\end{align}
Plugging this into Eq.~(\ref{Eq:Sol of Phi_n(x)-closed}) we find
\begin{align}
\tilde{\varphi}_n(x_0)=\frac{\sqrt{2}}{\sqrt{1+\chi_s+x_0\tan^2(\omega_nx_0)+(1-x_0)\tan^2[\omega_n(1-x_0)]}}
\label{Eq:HermEigMode@x0-Simplified}
\end{align}
\end{widetext}
Equations~(\ref{Eq:HermEigFreq-Simplified}) and (\ref{Eq:HermEigMode@x0-Simplified}) provide the asymptotic behavior of $\omega_n$, $\tilde{\varphi}_n(x_0)$ and $g_n$ for a general $x_0$. 
\section{Characteristic function $\large D_j(s)$ and its convergence}
\label{App:Char Func Dj(s)}
In this section we derive the expression for the characteristic function $D_j(s)$ and compare its convergence in number of resonator modes with and without the modification we found for $g_n$. 

Consider the Heisenberg-Langevin equations of motion corresponding to Hamiltonian~(\ref{Eq:2nd quantized closed H}) in the linear regime ($\epsilon=0$) for $\hat{\mathcal{X}}_{j,n}(t)$ as
\begin{subequations}
\begin{align}
&\left(d_t^2+\omega_j^2\right)\hat{\mathcal{X}}_j(t)=-\sum\limits_{n}2g_n\omega_n\hat{\mathcal{X}}_n(t),
\label{Eq:Heis Eq of Xj}\\
&\left(d_t^2+2\kappa_n d_t+\omega_n^2\right)\hat{\mathcal{X}}_n(t)=-2g_n\omega_j\hat{\mathcal{X}}_j(t)-\hat{f}_n(t),
\label{Eq:Heis Eq of Xn}
\end{align}
\end{subequations}
where $\kappa_n$ and $\hat{f}_n$ are the decay rate and noise operator coming from coupling to the waveguide degrees of freedom \cite{Senitzky_Dissipation_1960}.

Equations~(\ref{Eq:Heis Eq of Xj}-\ref{Eq:Heis Eq of Xn}) are linear constant coefficient ODEs and can be solved exactly via the unilateral Laplace transform 
\begin{align}
\tilde{h}(s)=\int_{0}^{\infty}dt h(t)e^{-st}.
\label{Eq:Def of Lap transform}
\end{align}

Taking the Laplace transform of Eqs.~(\ref{Eq:Heis Eq of Xj}-\ref{Eq:Heis Eq of Xn}) we obtain
\begin{subequations}
\begin{align}
\begin{split}
\left(s^2+\omega_j^2\right)\hat{\tilde{\mathcal{X}}}_j(s)+\sum\limits_n2g_n\omega_n \hat{\tilde{\mathcal{X}}}_n(s)=\\
s\hat{\mathcal{X}}_j(0)+\hat{\dot{\mathcal{X}}}_j(0),
\end{split}
\label{Eq:Laplace Eq of Xj}
\end{align}
\begin{align}
\begin{split}
\left(s^2+2\kappa_ns+\omega_n^2\right)\hat{\tilde{\mathcal{X}}}_n(s)+2g_n\omega_j\hat{\tilde{\mathcal{X}}}_j(s)=\\
(s+2\kappa_n)\hat{\mathcal{X}}_n(0)+\hat{\dot{\mathcal{X}}}_n(0)+\hat{\tilde{f}}(s).
\end{split}
\label{Eq:Laplace Eq of Xn}
\end{align}
\end{subequations}
The solution for $\hat{\tilde{\mathcal{X}}}_j(s)$ then reads
\begin{align}
\hat{\tilde{\mathcal{X}}}_j(s)=\frac{\hat{N}_j(s)}{D_j(s)},
\label{Eq:Laplace Sol of Xj}
\end{align}
where the numerator
\begin{align}
\begin{split}
\hat{N}_j(s)&=s\hat{\mathcal{X}}_j(0)+\hat{\dot{\mathcal{X}}}_j(0)\\
&-\sum\limits_n\frac{2g_n\omega_n\left[(s+2\kappa_n)\hat{\mathcal{X}}_n(0)+\hat{\dot{\mathcal{X}}}_n(0)-\hat{\tilde{f}}_n(s)\right]}{s^2+2\kappa_n s+\omega_n^2},
\end{split}
\label{Eq:Def of Nj(s)}
\end{align}
contains the operator initial conditions and the denominator
\begin{align}
D_j(s)\equiv s^2+\omega_j^2-\sum\limits_n\frac{4g_n^2\omega_j\omega_n}{s^2+2\kappa_n s+\omega_n^2}.
\label{Eq:Def of Dj(s)}
\end{align}
is the characteristic function whose roots give the hybridized poles of the full system. Therefore, we can represent $D_j(s)$ as
\begin{align}
D_j(s)=(s-p_j)(s-p_j^*)\prod\limits_n\frac{(s-p_n)(s-p_n^*)}{(s-z_n)(s-z_n^*)},
\label{Eq:Formal rep of Dj(s)}
\end{align}
where $p_{j,n}\equiv -\alpha_{j,n}-i\beta_{j,n}$ stand for the transmon-like and the $n$th resonator-like poles, respectively. Furthermore, $z_n \equiv-\kappa_n-i\sqrt{\omega_n^2-\kappa_n^2}$ is the $nth$ \textit{bare} non-Hermitian resonator mode. The notation ($p$ for poles and $z$ for zeros) is chosen based on $1/D_j(s)$ that appears in the Laplace solution~(\ref{Eq:Laplace Sol of Xj}). 

In order to compute the hybridized poles in practice, we need to truncate the number of resonator modes in $D_j(s)$. This truncation is only justified if the function $D_j(s)$ converges as we include more and more modes. First, note that without the correction give by $\chi_s$ this sum is divergent, since $g_n\thicksim \sqrt{\omega}_n\thicksim \sqrt{n}$ and for a fixed s we obtain
\begin{align}
\frac{4g_n^2\omega_j\omega_n}{s^2+2\kappa_n s+\omega_n^2}\thicksim \frac{\omega_n^2}{\omega_n^2}\thicksim 1.
\end{align}
Hence, the series in divergent. On the other hand, we found that for a non-zero $\chi_s$, $g_n\thicksim 1/\sqrt{\omega_n}\thicksim 1/\sqrt{n}$. Therefore we find
\begin{align}
\frac{4g_n^2\omega_j\omega_n}{s^2+2\kappa_n s+\omega_n^2}\thicksim \frac{1}{\omega_n^2}\thicksim \frac{1}{n^2},
\label{Eq:Convergence of Dj(s)}
\end{align}
and the series becomes convergent. In writing Eq.~(\ref{Eq:Convergence of Dj(s)}), we used the fact that $\omega_n\thicksim n$ and $\kappa_n$ has a sublinear asymptotic behavior found numerically.
\section{Divergence in the Wigner-Weisskopf theory of spontaneous emission}
Divergence of the Purcell decay rate appears in other frameworks besides the dispersive limit Jaynes-Cummings model as well. In this appendix, we show that the spontaneous decay rate of a qubit coupled to continuum of modes is also \textit{divergent}, unless the gauge invariance of the interaction is incorporated as presented in this manuscript. The impression of an (erroneous) finite decay rate in free space goes back to Wigner and Weisskopf's original work on spontaneous atomic decay, which implicitly makes a Markov approximation (See Sec.~$6.3$ of \cite{Scully_Quantum_1997}). We emphasize that employing the Markov approximation always yields a finite value for the decay rate regardless of the form of spectral function for electromagnetic background. 

To see this explicitly, we go over the Wigner-Weisskopf theory of spontaneous emission for a two-level system coupled to a continuum of modes inside an infinitely long 1D medium.  In interaction picture, the Hamiltonian reads 
\begin{align}
\hat{\mathcal{H}}_I=\sum\limits_{k}\hbar\left[g_{k}^*(x_0)\hat{\sigma}^+\hat{a}_{k}e^{i(\omega_j-\omega_{k})t}+H.c.\right],
\label{Eq:WW-Def of H_I}
\end{align}
which conserves the total number of excitations 
\begin{align}
\hat{N}\equiv \hat{\sigma}^+\hat{\sigma}^-+\sum\limits_{\vec{k}}\hat{a}_{\vec{k}}^{\dag}\hat{a}_{\vec{k}}.
\label{Eq:WW-Ansatz for Psi}
\end{align} 
As a result, a number conserving Ansatz for the wavefunction can be written as
\begin{align}
\ket{\Psi(t)}=c_e(t)\ket{e,0}+\sum\limits_{k}c_{g,k}(t)\ket{g,1_{k}},
\end{align}
where there is either no photon in the cavity and the qubit is in excited state $\ket{e}$, or there is a photon at frequency $\omega_k$ with qubit in the ground state $\ket{g}$. By solving the Schrodinger equation we obtain the time evolution of the unknown probability amplitudes $c_e(t)$ and $c_{g,k}(t)$. Combining these equations yields an effective equation for $c_e(t)$ as
\begin{align}
\dot{c}_e(t)=-\int_{0}^{t}dt'\mathcal{K}(t-t')c_e(t'),
\label{Eq:WW-Voltera Eq for Ce(t) 1}
\end{align}
where the memory Kernel $\mathcal{K}(\tau)$ is given by
\begin{align}
\mathcal{K}(\tau)\equiv\sum\limits_{k}|g_{k}(x_0)|^2e^{i(\omega_j-\omega_{k})t}.
\label{Eq:WW-Def of K(tau) 1}
\end{align}
Next, we replace the expression for $g_k(x_0)$, derived in Sec.~\ref{App:MultiJC}, as
\begin{align}
|g_{k}(x_0)|^2=\frac{\gamma\chi_s}{4}\omega_j\omega_k|\tilde{\varphi}_k(x_0)|^2.
\label{Eq:WW-Exp for g_k}
\end{align}
Note that without respecting the gauge symmetry of interaction $|\tilde{\varphi}_k(x_0)|=\mathcal{N}(x_0)$ is $k$-independent. Moreover, the sum over $k$ can be replaced as
\begin{align}
\sum\limits_{k}\rightarrow \frac{L}{2\pi}\int_{0}^{\infty}dk=\frac{L}{2\pi v_p}\int_{0}^{\infty}d\omega_{k},
\label{Eq:WW-Disc To Cont sum}
\end{align}
for a continuum of modes, where $v_p$ is the phase velocity of the medium. Inserting Eqs.~(\ref{Eq:WW-Exp for g_k}) and~(\ref{Eq:WW-Disc To Cont sum}) into the effective Eq.~(\ref{Eq:WW-Voltera Eq for Ce(t) 1}) we obtain
\begin{align}
\begin{split}
\dot{c}_e(t)&=-\frac{1}{2\pi}\frac{\gamma\chi_s\omega_j \mathcal{N}^2(x_0) L}{4v_p}\\
&\times\int_{0}^{\infty}d\omega_k \omega_k \int_{0}^{t}dt' e^{i(\omega_j-\omega_k)(t-t')}c_e(t')
\end{split}
\label{Eq:WW-Voltera Eq for Ce(t) 2}
\end{align}
Importantly, the integral over $\omega_k$ in Eq.~(\ref{Eq:WW-Voltera Eq for Ce(t) 2}) does not converge since the integrand grows unbounded as $\omega_k\to\infty$. To resolve this, Wigner and Weisskopf assumed that the dominant contribution comes from those modes of continuum whose frequency are close to the qubit frequency. Therefore, the factor $\omega_k$ can be replaced by $\omega_j$ and by extending the lower limit of integral over $\omega_k$ to $-\infty$ we can use the identity
\begin{align}
\int_{-\infty}^{+\infty}d\omega_{k}e^{i(\omega_j-\omega_{k})(t-t')}=2\pi\delta(t-t'),
\end{align}
to arrive at a \textit{finite} value for the spontaneous decay as
\begin{subequations}
\begin{align}
&\dot{c}_e(t)\approx-\frac{\Gamma_{sp}}{2}c_e(t),\\
&\Gamma_{sp}\equiv \frac{\gamma\chi_s\omega_j^2 \mathcal{N}^2(x_0) L}{2v_p}.
\end{align}
\end{subequations}
It is worth mentioning that using Markov approximation, one always obtains a finite expression for the spontaneous decay rate regardless of the form for the spectral function. This happens because instead of integrating over the entire frequency span, the Markov approximation picks a small window around qubit frequency.

Next, we show how our natural high frequency cut-off for light-matter coupling resolves the divergence of Wigner-Weisskopf theory. First, note that applying Markov approximation is indeed unnecessary, since the Volterra Eq.~(\ref{Eq:WW-Voltera Eq for Ce(t) 1}) with the memory kernel 
\begin{align}
\mathcal{K}(\tau)=\frac{1}{2\pi}\frac{\gamma\chi_s\omega_jL}{4v_p}\int_0^{\infty}d\omega_k \omega_k|\tilde{\varphi}_k(x_0)|^2e^{i(\omega_j-\omega_k)\tau},
\label{Eq:WW-Def of K(tau) 2}
\end{align}
has an exact solution in Laplace domain as
\begin{align}
\tilde{c}_e(s)=\frac{c_e(0)}{s+\tilde{\mathcal{K}}(s)},
\end{align}
where $\tilde{\mathcal{K}}(s)\equiv \int_{0}^{\infty}d\tau\mathcal{K}(\tau)e^{-s\tau}$ is the Laplace transform and is found as
\begin{align}
\tilde{\mathcal{K}}(s)=\frac{1}{2\pi}\frac{\gamma\chi_s\omega_jL}{4v_p}\int_0^{\infty}d\omega_k \frac{\omega_k|\tilde{\varphi}_k(x_0)|^2}{s+i(\omega_k-\omega_j)}.
\label{Eq:WW-tilde(K)(s)}
\end{align}
Second, when the gauge-invariance of the interaction is incorporated, the mode amplitude is frequency dependent that experiences a high frequency suppression as
\begin{align}
|\tilde{\varphi}_k(x_0)|\sim\frac{1}{\omega_k}.
\label{Eq:WW-Def of tilde(phi)_k(x_0)}
\end{align}
Replacing Eq.~(\ref{Eq:WW-Def of tilde(phi)_k(x_0)}) into expression~(\ref{Eq:WW-tilde(K)(s)}) for $\tilde{\mathcal{K}}(s)$ we obtain
\begin{align}
\tilde{\mathcal{K}}(s)\propto\int d\omega_k\frac{1}{\omega_k[s+i(\omega_k-\omega_j)]}.
\end{align}
Interestingly, with the corrected expression for the eigenmodes, the integrand behaves like $1/\omega_k^2$ at $\omega_k\to \infty$, and as a result the integral converges. Otherwise, the integrand behaves like a constant at $\omega_k\to \infty$ and the result is divergent.
\color{black}
\bibliographystyle{apsrev4-1}
\bibliography{MulticQED}
\end{document}